\documentclass[reprint,
aps,nofootinbib,
longbibliography,
pra]{revtex4-1}
\usepackage{amssymb}
\usepackage{eurosym}
\usepackage{amsmath}
\usepackage{graphicx}
\usepackage{dcolumn}
\usepackage{bm}
\usepackage{txfonts}
\usepackage{color}
\usepackage[unicode,breaklinks]{hyperref}

\hypersetup{
    unicode=true,
    a4paper=true,
    plainpages=false,
    pdftitle={Splitting of two-component solitons from collisions with narrow potential barriers},
    pdfauthor={C. L. Grimshaw, S. A. Gardiner, B. A. Malomed},
    pdfsubject={Splitting of two-component solitons from collisions with narrow potential barriers},
    colorlinks=true,
    linkcolor=blue,
    citecolor=blue,
    filecolor=black,
    urlcolor=blue
}
\begin{document}

\title{Splitting of two-component solitary waves from collisions with narrow
potential barriers}
\author{Callum L. Grimshaw and Simon A. Gardiner}
\author{Boris A. Malomed}
\affiliation{Joint Quantum Centre (JQC) Durham--Newcastle, Department of Physics, Durham University, Durham, DH1 3LE,
United Kingdom}
\affiliation{Department of Physical Electronics, School of Electrical
Engineering, Faculty of Engineering, and Center for Light--Matter
Interaction, Tel Aviv University, Tel Aviv 69978, Israel}
\date{\today }

\begin{abstract}
We consider the interaction of two-component bright-bright solitons with a
narrow potential barrier (splitter) in the framework of a system of two
Gross-Pitaevskii (nonlinear-Schr\"{o}dinger) equations modeling a binary
Bose-Einstein condensate, with self-attraction in each component and
cross-attraction between them. The objective is to study splitting of
composite solitons, which may be used in the design of two-component
soliton-interferometer schemes. We produce approximate analytic results,
assuming a weak barrier and applying the perturbation theory in the limit in
which the system is integrable and the solitary waves may be considered as
exact solitons. We do this in the case of negligible interspecies
interactions, and also when the nonlinearities are strongly asymmetric,
allowing one to neglect the self-interaction in one of the species. Then, we
use systematic simulations to study the transmissions of both components in
regions outside these approximations and, in particular, to compare
numerical results with their analytical countertparts. We concluded that
there is an appreciable parameter range where one component is almost
entirely transmitted through the barrier, while the other one is reflected.
Excitation of internal vibrations in the passing and rebounding solitons is
explored too, with a conclusion that it is weak in the regime of
high-quality splitting.
\end{abstract}

\maketitle

\section{Introduction \label{Sec:Introduction}}

Solitons, or more broadly solitary waves, are manifest in a broad range of
physical settings \cite{Peyrard,KA}, including, in particular, nonlinear
matter waves in atomic Bose--Einstein condensates(BECs) \cite
{Randy,SP,khaykovich,strathclyde}. In the mean-field approximation, the
commonly adopted dynamical model of a BEC is based on the Gross--Pitaevskii
equation (GPE) for a single-component condensate, and a system of coupled
GPEs for binary (two-component) mixtures \cite{Pit}. One of the potential
applications of matter-wave solitons is their use in the design of
interferometers, in which an incident soliton splits into two fragments upon
hitting a narrow potential barrier, followed by recombination of the
fragments after rebounding from the confining potential. An object to be
probed by the interferometer is placed as an obstacle through which one
fragment will pass, which will affect the outcome of the recombination \cite
{si1,exper,wales}. Soliton interferometers have been elaborated
theoretically in various configurations \cite
{interf1,interf2,interf3,interf4,interf5,interf6,interf7,interf8,interf10,interf11}
(including the case when the splitter is inserted as a localized
self-repulsive nonlinearity, or its combination with the usual potential
barrier \cite{HS}) and realized in experiment \cite{exper}. Interactions of
matter-wave solitons with local potentials have also been studied in other
contexts, such as an analytical treatment of the collisions \cite
{dfanalytics1}, rebound from potential wells \cite{Brand1,Brand2}, dynamics
of solitons in a dipolar BEC \cite{interf2}, and probing effects of
interparticle interactions on tunneling \cite{tunneling1,tunneling2}.
However, the splitting of a fundamental soliton by a linear and/or nonlinear
potential barrier implies, in a sense, the application of a
\textquotedblleft brute force\textquotedblright\ to the soliton, as its
intrinsic structure does not resonate with the action of the splitter. A
more natural option, which was elaborated recently, is fission of a
2-soliton (breather) into its fundamental-soliton constituents \cite{Oleks}
(see also Ref. \cite{Vanja}), with the amplitude ratio close to the natural
value, $3:1$ \cite{SY} (see also \cite{billam_1} for a similarly motivated
protocol involving a laser pulse in combination with control of the
scattering length). These settings may also be realized in the context of
optics, with GPE replaced by the nonlinear Schr\"{o}dinger equation (NLSE)
for the spatial-domain propagation of light in planar waveguides \cite{KA}.

In this work, we aim to elaborate another natural scheme for the splitting,
when an incident two-component soliton, governed by a pair of nonlinearly
coupled GPEs, hits a narrow splitting barrier. The situation under
consideration is one with equal atomic masses and equal negative scattering
lengths in the two components, and attractive interaction between the
components, while there is no linear coupling (interconversion) between them
(interconversion would make splitting of a composite soliton into
single-component ones impossible). We note that replacing time in the
coupled GPEs by the propagation distance, $z$, this model also applies to
bimodal light propagation in a Kerr-nonlinear waveguide with transverse
coordinate $x$, where $\psi _{1}$ and $\psi _{2}$ are amplitudes of two
components of the electromagnetic wave, corresponding to different carrier
wavelengths \cite{KA}, and where the potential represents transverse
modulation of the refractive index. However, in this case the strength of
the cross-interaction can only take a single value [$g=2$, in terms of the
notation adopted below in Eqs. (\ref{1}) and (\ref{2})], as there is no
straightforward optical counterpart to the Feshbach-resonance technique.
Alternatively, if $\psi _{1}$ and $\psi _{2}$ represent amplitudes of two
waves with mutually orthogonal linear polarizations, the relevant value is $
g=2/3$, provided that rapidly oscillating four-wave-mixing terms may be
neglected \cite{KA}.

As mentioned above, previous works have addressed collisions of
single-component solitons with potential barriers, represented by an ideal $
\delta $-function or a narrow Gaussian potential barrier, aiming to identify
outcomes of the collisions as functions of the velocity of the incoming
soliton and the barrier's height and width \cite{si1,dfanalytics1}. Dynamics
of two-component solitons has been studied with regard to their intrinsic
vibrations in free space \cite{of2}, as well as collisions between two
solitons in the presence of a narrow Gaussian barrier added to the Manakov's
system with equal coefficients of the self- and cross-attraction \cite
{bin_int}, and also the scattering of dark-bright solitons by impurities
\cite{dbscat}. The main objective of the present work is to identify a
parameter region in which the collision of a composite solitary wave with
the barrier effectively splits it into single-component constituents. The
primary control parameters that we consider are the relative norm of the
components, defined as parameter $f$, the velocity of the incident soliton,
the strength of the barrier $\varepsilon $ [see Eq.~(\ref{sigma})], and the
relative strength of the interspecies attraction $g$ [see Eqs.~(\ref{1}) and
(\ref{2})]. We first report approximate analytical results, obtained for the
system with a weak barrier, in Section II. We then summarize results of
systematic numerical simulations of the collisions in Section III. We
compare analytical results to their numerical counterparts in Section III,
and conclude the paper by Section IV. Some technical details are presented
in Appendices A and B.

\section{The system \label{Sec:System}}

We consider a binary BEC system, with two components corresponding to
different internal states of the same atomic species, and collisions
dominated by the $s$-wave scattering. We model the system by two coupled
GPEs, assuming, as usual, that the mean-field wave functions of the two
components are radially confined by a tight trapping potential in the
transverse $\left( y,z\right) $ plane, and weakly confined in the axial ($x$
) direction, if at all. In addition, we assume that an off-resonant sheet of
light, propagating perpendicular to the axial direction, with peak beam
strength $E_{\mathrm{B}}$ and axial width $x_{r}$ (defined at the
relative amplitude level $1/e^{2}$), creates a barrier potential for
both components, centered at $x=0$ \cite{marchant,wales}. We assume that the
transverse and barrier potentials are insensitive to the internal atomic
state, allowing the coupled GPEs to take the form of
\begin{subequations}
\begin{align}
\mathrm{i}\hbar \frac{\partial }{\partial t}\Psi _{1}(\mathbf{r})=& \left[ -
\frac{\hbar ^{2}\nabla ^{2}}{2m}+V(x)+\frac{1}{2}m\omega _{r}^{2}\left(
y^{2}+z^{2}\right) \right] \Psi _{1}(\mathbf{r})  \notag \\
& +\frac{4\pi \hbar ^{2}N}{m}\left[ a_{11}|\Psi _{1}(\mathbf{r}
)|^{2}+a_{12}|\Psi _{2}(\mathbf{r})|^{2}\right] \Psi _{1}(\mathbf{r}),
\label{1_3d} \\
\mathrm{i}\hbar \frac{\partial }{\partial t}\Psi _{2}(\mathbf{r})=& \left[ -
\frac{\hbar ^{2}\nabla ^{2}}{2m}+V(x)+\frac{1}{2}m\omega _{r}^{2}\left(
y^{2}+z^{2}\right) \right] \Psi _{2}(\mathbf{r})  \notag \\
& +\frac{4\pi \hbar ^{2}N}{m}\left[ a_{22}|\Psi _{2}(\mathbf{r}
)|^{2}+a_{12}|\Psi _{1}(\mathbf{r})|^{2}\right] \Psi _{2}(\mathbf{r}),
\label{2_3d}
\end{align}
where $m$ is the atomic mass, $a_{11}$, $a_{22}$, and $a_{12}$ are the
intra- and inter-species $s$-wave scattering lengths, $V(x)=E_{\mathrm{B}}
\mathrm{e}^{-2x^{2}/x_{r}^{2}}+m\omega _{\mathrm{T}}^{2}x^{2}/2$ is, as said
above, the combination of the barrier potentials and a weak axial trapping
one, $\omega _{\mathrm{T}}$ and $\omega _{r}$ are axial and radial trapping
frequencies, and $N$ is the total number of particles. Equations (\ref{1_3d}
) and (\ref{2_3d}) are supplemented by the normalization convention
\end{subequations}
\begin{equation}
\int \mathop{\mathrm{d}\mathbf{r}}|\Psi _{1}(\mathbf{r})|^{2}=f,\int 
\mathop{\mathrm{d}\mathbf{r}}|\Psi _{2}(\mathbf{r})|^{2}=1-f,  \label{f}
\end{equation}
so that
\begin{equation}
\int \mathop{\mathrm{d}\mathbf{r}}\left[ |\Psi _{1}(\mathbf{r})|^{2}+|\Psi
_{2}(\mathbf{r})|^{2}\right] =1,
\end{equation}
hence the numbers of particles in the two components are $N_{1}=fN$ and $
N_{2}=(1-f)N$.

Strong radial confinement then permits us to assume the usual Gaussian
ansatz $\phi (y,z)=(m\omega _{r}/\pi \hbar )^{1/2}\exp (-m\omega
_{r}[y^{2}+z^{2}]/2\hbar )$ for the radial degrees of freedom of the
condensate wavefunctions $\Psi _{1}(\mathbf{r})$, $\Psi _{2}(\mathbf{r})$.
We integrate over the transverse plane (with coordinates $y$ and $z$),
define $g_{11}\equiv 2\hbar \omega _{r}a_{11}$, and express the result in
terms of notation with unit coordinate $\hbar ^{2}/m|g_{11}|N$, unit time $
\hbar ^{3}/m(g_{11}N)^{2}$, and unit energy $m(g_{11}N/\hbar )^{2}$ [which
implies unit velocity $|g_{11}|N/\hbar $, and that, after the integration
over the transverse plane, we multiply the condensate wavefunctions by $
\hbar /(m|g_{11}|N)^{1/2}$ to render them dimensionless].\footnote{
This can be thought of heuristically as notation with $\hbar =m=g_{11}N=1$.}
As a result, the three-dimensional (3D) system of Eqs. (\ref{1_3d}) and (\ref
{2_3d}) is reduced to the 1D form:
\begin{subequations}
\label{1_1dand2_1d}
\begin{align}
\mathrm{i}\frac{\partial }{\partial t}\psi _{1}(x)=& \left[ -\frac{1}{2}
\frac{\partial ^{2}}{\partial x^{2}}+\frac{1}{2}\omega
_{x}^{2}x^{2}+\varepsilon \eta (x,\sigma )\right] \psi _{1}(x)  \notag \\
& -\left[ |\psi _{1}(x)|^{2}+g|\psi _{2}(x)|^{2}\right] \psi _{1}(x),
\label{1_1d} \\
\mathrm{i}\frac{\partial }{\partial t}\psi _{2}(x)=& \left[ -\frac{1}{2}
\frac{\partial ^{2}}{\partial x^{2}}+\frac{1}{2}\omega
_{x}^{2}x^{2}+\varepsilon \eta (x,\sigma )\right] \psi _{2}(x)  \notag \\
& -\left[ g^{\prime }|\psi _{2}(x)|^{2}+g|\psi _{1}(x)|^{2}\right] \psi
_{2}(x),  \label{2_1d}
\end{align}
where $g=a_{12}/a_{11}$, $g^{\prime }=a_{22}/a_{11}$, $\omega _{x}=\omega _{
\mathrm{T}}\hbar ^{3}/m|g_{11}|^{2}N^{2}$, $\varepsilon =E_{\mathrm{B}
}x_{r}(\pi /2)^{1/2}/m^{2}|g_{11}|^{3}N^{3}$, $\sigma =x_{r}/2\hbar
^{2}/m|g_{11}|N$, and
\end{subequations}
\begin{equation}
\eta (x,\sigma )=\frac{1}{\sqrt{2\pi }\sigma }\exp (-x^{2}/2\sigma ^{2}),
\label{sigma}
\end{equation}
such that $\lim_{\sigma \rightarrow 0}\eta (x,\sigma )=\delta (x)$, and we
have assumed all the scattering lengths to be negative (i.e., that all the
interactions are attractive). Note that, as mentioned above, the relative
strength of the cross-attraction between the components, $g$, can be
effectively adjusted by means of the Feshbach-resonance technique \cite
{feshbachresonance,Yb}, and $\varepsilon >0$ is the strength of the
splitting barrier. Direct control of properties of binary BECs has been
demonstrated in Ref. \cite{hetero_fr} for a heteronuclear BEC, and in Ref.
\cite{homo_fr} for a BEC composed of different hyperfine states of the same
species, where the interspecies interaction was varied to probe the
miscibility-immiscibility transition. Although it has become a fairly
standard technique to control the scattering length in BEC systems,
conventionally using magnetic Feshbach resonances, there are limitations as
to what can be achieved in multicomponent systems. For instance, when
exploiting a magnetic Feshbach resonance, in principle all of the scattering
lengths depend on the value of the applied magnetic field and therefore
cannot be varied independently. Hence, the Feshbach resonance technique must
be utilized in such a way that the three scattering lengths (in the
two-component system studied here) are brought as close as possible to their
desired values. For the numerical results presented in Section III we fix $
g^{\prime }=1$ in Eq. (\ref{2_1d}) (i.e., $a_{11}=a_{22}$) and vary $g$. We
point out that in the particular case of $g=0$, in the GPE system considered
in this paper, the value of $g^{\prime }$ has a role equivalent to that of $f
$ [see Eq. (\ref{f})], in that it defines the relative self-interaction
strength of the two condensate components. Furthermore, there may be regimes
where $g$, for example, can be controlled essentially independently, keeping
the intraspecies scattering lengths close to their background values over
relevant values of the magnetic field and $a_{11}\approx a_{22}$. The use of
the optical Feshbach-resonance techniques may also be considered, although
in this case one would need to accept the large loss rates that typically
accompany the optically-induced resonance \cite
{exp_opt_feshbach1,exp_opt_feshbach2}. Still another possibility is the use
of laser-assisted magnetic Feshbach resonances \cite{opt_ass_feshbach}. To
some extent, one additionally has a choice between different atomic species,
with different background values of the scattering lengths, and their
different dependences on the applied fields.

In this paper we always assume $(1-f)g^{\prime }\geq f$, i.e., $f\leq
g^{\prime }/(1+g^{\prime })$, which effectively defines component 2 as the one
with the largest intraspecies mean-field attraction; put in terms of
particle numbers and scattering lengths, this means that $N_{1}a_{11}\leq
N_{2}a_{22}$. In the case of $g^{\prime }=1$ (i.e., $a_{22}=a_{11}$), this
assumption reduces to $f\leq 1/2$ (i.e., $N_{1}\leq N_{2}$); and in the case
of $f=1/2$ (i.e., $N_{1}=N_{2}$), it reduces to $g^{\prime }\geq 1$ ($
a_{22}\geq a_{11}$). In our analytical treatment we choose, in general, to
keep both $f$ and $g^{\prime }$ as separate parameters that may be
individually varied. Then, variation of $f$ corresponds to relative changes
in the numbers of particles in the two components, and the variation of $
g^{\prime }$ (and $g$) correspond to changes in the (negative) scattering
lengths. This is an appropriate representation of what may be altered in the
experimental work with BEC, although the basic results presented in the next
section may be readily understood too in the simpler limit case of $g^{\prime }=1$.

\section{Analytical considerations \label{Sec:AnalyticalApproach}}

\subsection{The simplified system \label{Sec:IdealisedSystem}}

To address an infinitesimally narrow barrier in free 1D space, we set $
\omega _{x}=0$ and $\sigma \rightarrow 0$ in Eqs.~(\ref{1_1dand2_1d}) and (
\ref{sigma}), which leads to a more analytically tractable form of the GPE
system:
\begin{subequations}
\label{1and2}
\begin{align}
\mathrm{i}\frac{\partial \psi _{1}}{\partial t}=& \left[ -\frac{1}{2}\frac{
\partial ^{2}}{\partial x^{2}}+\varepsilon \delta (x)-|\psi _{1}|^{2}-g|\psi
_{2}|^{2}\right] \psi _{1},  \label{1} \\
\mathrm{i}\frac{\partial \psi _{2}}{\partial t}=& \left[ -\frac{1}{2}\frac{
\partial ^{2}}{\partial x^{2}}+\varepsilon \delta (x)-g^{\prime }|\psi
_{2}|^{2}-g|\psi _{1}|^{2}\right] \psi _{2},  \label{2}
\end{align}%
which we use in this section. The system can be derived from the Hamiltonian
(conserved energy of the mean-field theory),
\end{subequations}
\begin{equation}
\begin{split}
H[\psi _{1},\psi _{1}^{\ast },\psi _{2},\psi _{2}^{\ast }]=& \int_{-\infty
}^{+\infty }\mathop{\mathrm{d}x}\Bigg[\frac{1}{2}\left( \left\vert \frac{
\partial \psi _{1}}{\partial x}\right\vert ^{2}+\left\vert \frac{\partial
\psi _{2}}{\partial x}\right\vert ^{2}\right) \\
& -\frac{1}{2}\left( |\psi _{1}|^{4}+g^{\prime }|\psi _{2}|^{4}+2g|\psi
_{1}|^{2}|\psi _{2}|^{2}\right) \\
& +\varepsilon \delta (x)\left( |\psi _{1}|^{2}+|\psi _{2}|^{2}\right) \Bigg] .
\end{split}
\label{H}
\end{equation}

Setting $\varepsilon =0$, two integrable cases of the system can be
identified: If $g=0$, Eqs.~(\ref{1and2}) reduce to a pair of uncoupled
integrable NLSEs; and, if $g=g^{\prime }=1$, Eqs.~(\ref{1and2}) constitute
the integrable Manakov system \cite{manakov1}. Note that the case of $g=-1$,
which corresponds to the repulsive interspecies interaction is also
integrable. It should also be noted that, even in the attractive system in
which the Manakov condition, $g=g^{\prime }=1$, does not hold,
two-component bright soliton solutions, with identical mode profiles of both
components proportional to $\mathop{\mathrm{sech}}(F[x-(x_{0}+vt)]/2)$ [the
same form as in Eqs.~(\ref{psigll1and2}), see below], exist if $
f+(1-f)g=(1-f)g^{\prime }+fg\equiv F$. The latter condition is tantamount to
setting $f=(g^{\prime }-g)/(g^{\prime }+1-2g)$, or, in terms of the numbers
of particles and (negative) scattering lengths,
\begin{equation}
\frac{N_{1}}{N}=\frac{a_{12}-a_{22}}{2a_{12}-a_{11}-a_{22}}.
\end{equation}

The availability of the integrable forms of the system for $\varepsilon =0$
provides a natural framework for application of the perturbation theory in
the case of relatively small $\varepsilon $.

\subsection{The limit of negligible interspecies interactions \label{Sec:LimitNegligible}}

Here we address Eqs.~(\ref{1and2}) with $g=0$ (no interspecies
interactions), with an input in the form of a composite soliton with
coinciding centers of both species, moving as $\xi (t)=x_{0}+vt$. Far from
the potential barrier, the composite soliton is built as
\begin{subequations}
\label{psigll1and2}
\begin{align}
\psi _{1}(x,t)=& \frac{1}{2}f\exp \left[ \mathrm{i}\left( vx-\mu
_{1}t\right) \right]  \notag \\
& \times \mathop{\mathrm{sech}}\left( \frac{1}{2}f\left[ x-\left(
x_{0}+vt\right) \right] \right) ,  \label{psigll1} \\
\psi _{2}(x,t)=& \frac{1}{2}(1-f)\sqrt{g^{\prime }}\exp \left[ \mathrm{i}
\left( vx-\mu _{1}t\right) \right]  \notag \\
& \times \mathop{\mathrm{sech}}\left( \frac{1}{2}\left( 1-f\right) g^{\prime
}\left[ x-\left( x_{0}+vt\right) \right] \right) ,  \label{psigll2}
\end{align}%
with chemical potentials
\end{subequations}
\begin{subequations}
\label{mu}
\begin{align}
\mu _{1}=& -f^{2}/8+v^{2}/2,  \label{mu1} \\
\mu _{2}=& -\left( 1-f\right) ^{2}g^{\prime 2}/8+v^{2}/2.  \label{mu2}
\end{align}

In the present case of $g=0$, the perturbation theory is based on a natural
conjecture that each component passes the barrier under the condition that
its center-of-mass kinetic energy exceeds the respective peak potential
energy, as given by Eq.~(\ref{U12}) in Appendix \ref{lowganalysis}.
Utilizing the solutions of Eqs.~(\ref{psigll1and2}), we have calculated the
respective energy terms from Eq.~(\ref{H}), as summarized in Appendix \ref{lowganalysis}, and used them to identify regimes in which the components do
or do not pass the barrier. Also, we have used these terms to estimate a
validity region of this approach. From Eqs.~(\ref{Pot1and2}) and (\ref{U12})
it follows that the condition for $\varepsilon $ to be sufficiently small is
\end{subequations}
\begin{equation}
\varepsilon \ll f/3.  \label{applicability}
\end{equation}

Combining Eqs.~(\ref{kinetic_energy}) and Eq.~(\ref{U12}) results in the
conditions
\begin{subequations}
\begin{align}
v^{2}>& \varepsilon f/2,  \label{pass1} \\
v^{2}>& \varepsilon (1-f)g^{\prime }/2,  \label{pass2}
\end{align}
for components 1 and 2, respectively, to be transmitted. We thus predict the
incident composite soliton to be split into a pair of a transmitted soliton
in component 1 and a reflected one in component 2 in the following interval of
velocities:
\end{subequations}
\begin{equation}
\sqrt{\varepsilon f/2}<|v|<\sqrt{\varepsilon (1-f)g^{\prime }/2}
\label{sep_condgll1}
\end{equation}%
[recall that we set $f\leq (1-f)g^{\prime }$]. We expect that the prediction
given by Eq. (\ref{sep_condgll1}) remains valid in the case of nonzero but
weak interspecies attraction, $g\ll 1$.

A similar approach can also be used in the case when the splitter is
nonlinear (corresponding to a steep increase in the interspecies scattering
length over a small spatial region). The analysis for the latter case is
produced in Appendix \ref{Sec:ExtensionNonlinear}.

\begin{figure}[t]
\includegraphics[width=\linewidth,trim=2cm 2cm 2cm
1.5cm,clip=true]{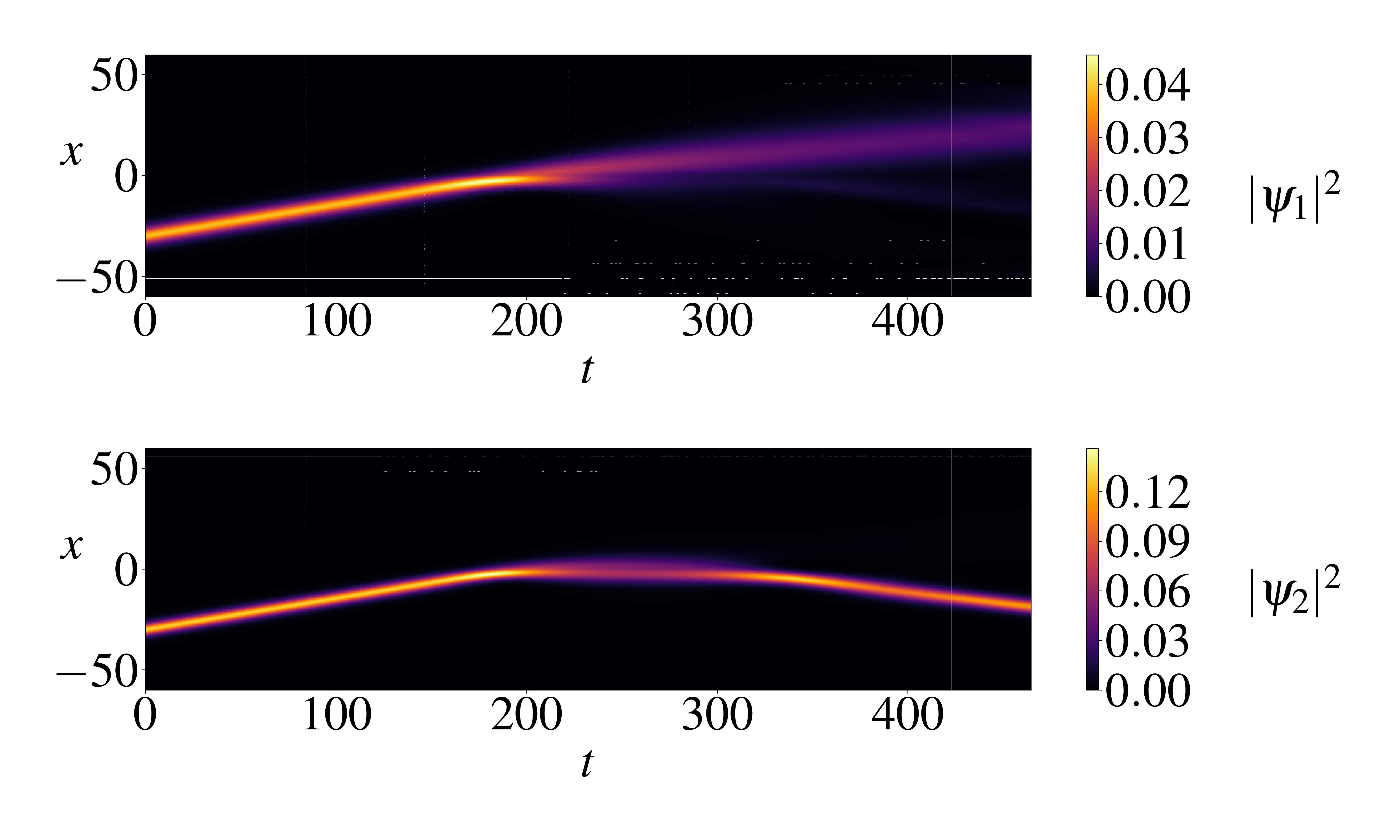}
\caption{The evolution of densities of the two components, which
demonstrates the splitting of the incident composite soliton into the
reflected heavier component and transmitted lighter component close to the
splitting-unsplitting boundary, at parameter values $\protect\sigma =0.4$, $\protect\varepsilon =0.07$, $f=0.3$, $g=0.2$, and $v=0.155$.}
\label{traj_sep}
\end{figure}

\begin{figure}[t]
\includegraphics[width=\linewidth,trim=2cm 2cm 2cm
1.5cm,clip=true]{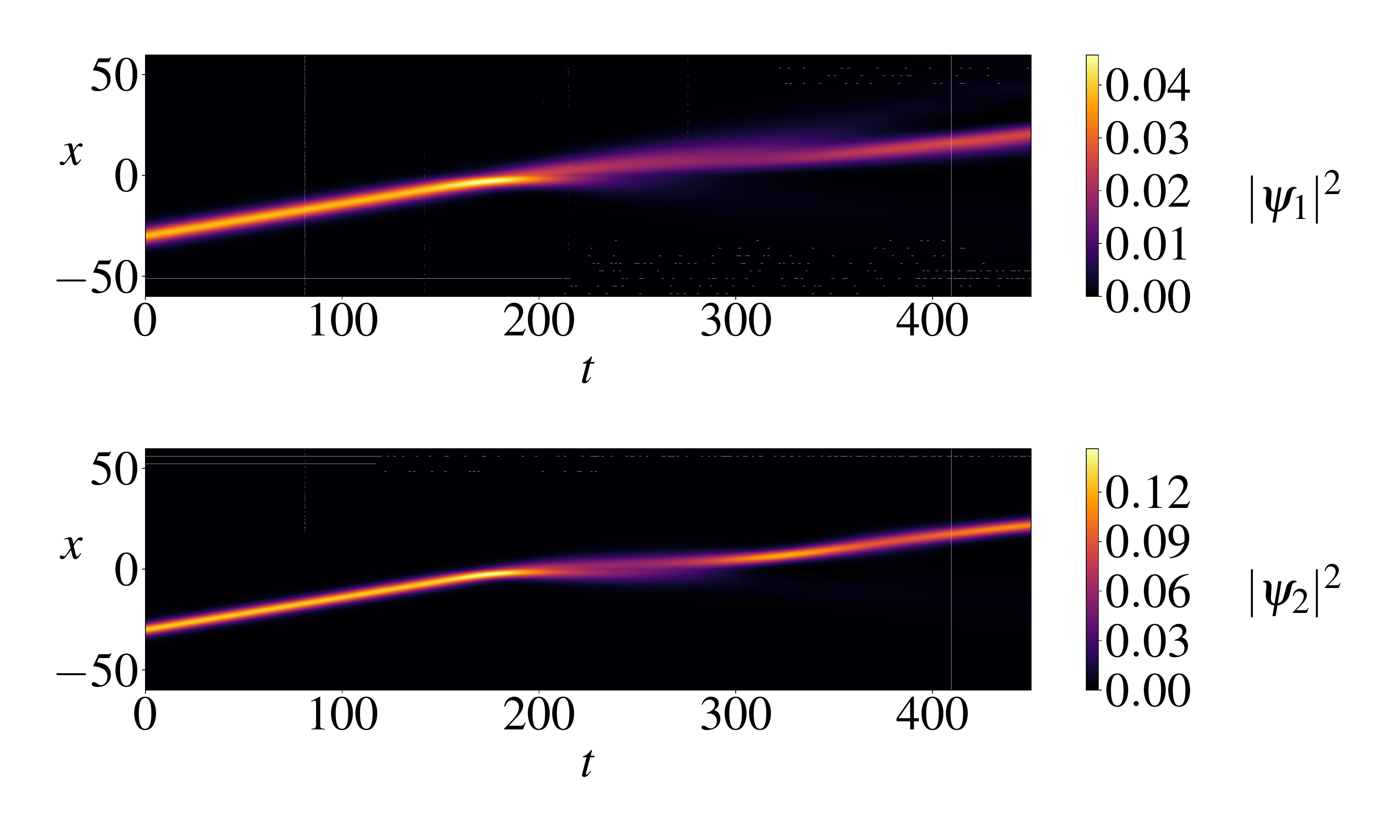}
\caption{The same as in Fig. \protect\ref{traj_sep}, but for the case when
the incident composite soliton passes the potential barrier without
splitting, close to the splitting threshold. Parameters are the same as in
Fig. \protect\ref{traj_sep}, except for a slightly larger value of the
collision velocity, $v=0.16$.}
\label{traj_nosep}
\end{figure}

\subsection{The limit of a strongly asymmetric two-component soliton \label{Sec:LimitStrongly}}

One can also carry out a perturbative analysis for small $\varepsilon $ in
the case when the intraspecies self-interaction of component 1 is much
weaker than its attractive interaction with component 2, i.e., $f\ll (1-f)g$
, and, accordingly, the intraspecies self-attraction of component 2 is much
stronger than its interaction with component 1, i.e., $fg\ll (1-f)g^{\prime
} $. These conditions are summarised as
\begin{equation}
\frac{f}{(1-f)}\ll g\ll \frac{g^{\prime }(1-f)}{f},
\label{asymmetric_conditions}
\end{equation}
which simplify to $f/(1-f)\ll g\ll (1-f)/f$ when $g^{\prime }=1$, and to $
1\ll g\ll g^{\prime }$ when $f=1-f=1/2$.

\begin{figure*}[t]
\includegraphics[width=\linewidth]{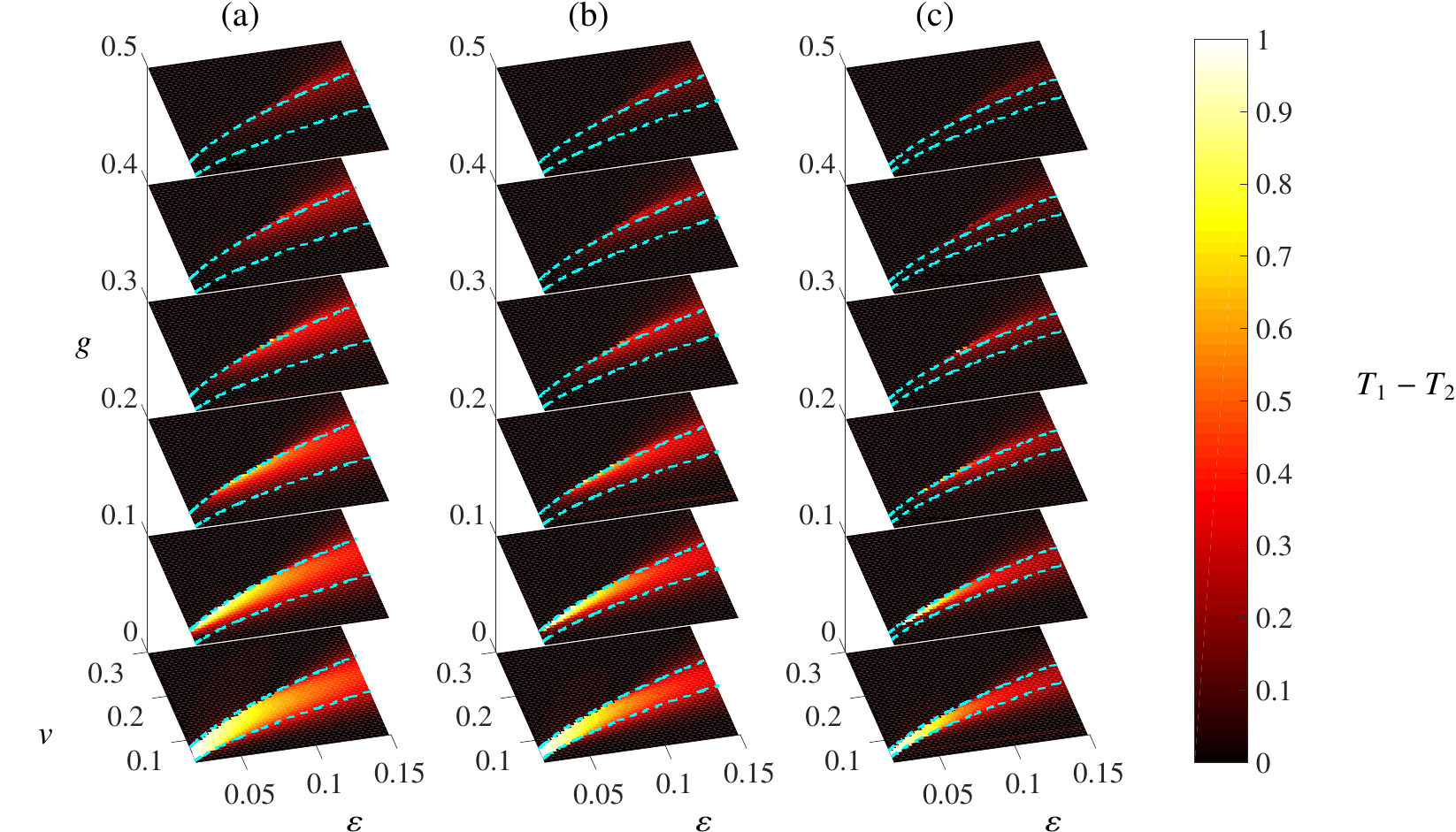}
\caption{The transmission difference between the two components, $T_{1}-T_{2}
$, as produced by simulations of Eqs. (\protect\ref{1}) and (\protect\ref{2}
), in the $(v,\protect\varepsilon )$ parameter plane (where $v$ is the
collision velocity and $\protect\varepsilon $ the strength of the Gaussian
potential barrier) at different values of $g$ and $f$ (the cross-attraction
strength and scaled population of the first component, respectively). The
value of $g$ increases along the vertical axis. (a) $f=0.3$, (b) $f=0.35$,
and (c) $f=0.4$. Here and in Fig. \protect\ref{fig_fv3d} the dashed lines
display boundaries of the splitting region, as analytically predicted by
Eq.~(\protect\ref{sep_condgll1}) in the limit of $g=0$.}
\label{fig_ev3d}
\end{figure*}

In the present case, component 2 of the incident mode is essentially the
usual NLSE bright soliton, as given by Eq.~(\ref{psigll2}) and Eq.~(\ref{mu2}),
while a solution for component 1 is sought for as
\begin{equation}
\psi _{1}(x,t)=\exp \left( \mathrm{i}\left[ vx-\left( \mu
_{1}^{(0)}+v^{2}/2\right) t\right] \right) u_{1}\left( x-\left[ x_{0}+vt
\right] \right) ,
\end{equation}
with $u_{1}$ determined by a stationary linear Schr\"{o}dinger equation:

\begin{equation}
\begin{split}
\mu _{1}^{(0)}u_{1}(X)=& -\frac{1}{2}\frac{d^{2}}{dX^{2}}u_{1}(X) \\
& -\frac{1}{4}(1-f)^{2}gg^{\prime }\mathop{\mathrm{sech}^{2}}\left[ \frac{1}{
2}(1-f)g^{\prime }X\right] u_{1}(X),
\end{split}
\label{linear}
\end{equation}%
where $X\equiv x-(x_{0}+vt)\equiv X$. Equation (\ref{linear}) describes a 1D
quantum particle in a P\"{o}schl--Teller potential, which, generally, can be
solved in terms of special functions.\footnote{%
This is made substantially simpler upon implementing the change of variable $
Y=(1-f)g^{\prime }X$.} The exact ground-state solution to Eq.~(\ref{linear})
and the corresponding eigenvalue are given by
\begin{subequations}
\begin{align}
u_{1}(X)=& A_{1}\left( \mathrm{sech}\left[ \frac{1}{2}(1-f)g^{\prime }X
\right] \right) ^{\alpha },  \label{a} \\
(\mu _{1}^{(0)})_{\mathrm{ground}}=& -\frac{1}{8}\alpha (1-f)g^{\prime 2},
\label{mu-GS}
\end{align}%
where
\end{subequations}
\begin{equation}
\alpha =\sqrt{\frac{1}{4}+\frac{2g}{g^{\prime }}}-\frac{1}{2}.  \label{alpha}
\end{equation}

Amplitude $A_{1}$ in Eq. (\ref{a}) is, by itself, arbitrary. However, as the
norm of component 1 is fixed to be $f$, $A_{1}$ is determined by condition
%\begin{equation}
\begin{equation}
f\equiv \int_{-\infty }^{+\infty }dX|u_{1}(X)|^{2}=A_{1}^{2}\frac{2\sqrt{\pi
}\mathop{\Gamma(\alpha)}}{(1-f)g^{\prime }\mathop{\Gamma \left( \alpha
+1/2\right)}},
%\end{equation}
\label{N1}
\end{equation}
where $\Gamma $ is the Gamma function.

Next, we follow the same approach as in the previous subsection, but with
the wave form of component 1 given by Eq.~(\ref{a}), to identify regimes in
which component 1 is transmitted (see Appendix \ref{Sec:DeterminingSplitting}
), while component 2 is reflected by the barrier. Thus, the condition for
the reflection of component 1 is obtained as
\begin{equation}
v^{2}>\frac{\varepsilon (1-f)g^{\prime }\mathop{\Gamma \left( \alpha
+1/2\right)}}{\sqrt{\pi }\mathop{\Gamma(\alpha)}}.
\end{equation}%
On the other hand, the condition for the reflection of component 2 remains,
in the first approximation, the same as given by Eq.~(\ref{pass2}). Hence,
this condition becomes $v^{2}<\varepsilon g^{\prime }(1-f)/2$, and there are
intervals of velocities defined by
\begin{equation}
\sqrt{\frac{\varepsilon (1-f)g^{\prime }\mathop{\Gamma( \alpha +1/2)}}{\sqrt{
\pi }\mathop{\Gamma(\alpha)}}}<|v|<\sqrt{\frac{\varepsilon (1-f)g^{\prime }}{
2}},  \label{g<1}
\end{equation}%
[cf.\ Eq.~(\ref{sep_condgll1})] in which the collision of the incident
composite soliton with the barrier leads to splitting, with component 1
transmitted and component 2 reflected. Note that when $\alpha =1$ [as
follows from Eq.~(\ref{alpha}), this happens at $g/g^{\prime }=1$] the
result is $\sqrt{\pi }\mathop{\Gamma(\alpha)}=2\mathop{\Gamma(\alpha + 1/2)}$
, and interval (\ref{g<1})\ shrinks to nil; this, in particular, applies for
the Manakov system, when $g=g^{\prime }=1$. For $g/g^{\prime }>1$ the
situation inverts, and there is a velocity interval, defined by transposing
the upper and lower bounds in Eq.~(\ref{g<1}), in which component 1 is
reflected and component 2 transmitted.

It is important to note that, in the limit case of $g^{\prime }=1$, this
regime is accessed just by adjusting the relative particle numbers in the
two components, with a much smaller population in component 1. Lastly, it is
relevant to mention that, while this consideration is entirely valid in the
framework of the mean-field theory, it may become irrelevant in the case of
a very small number of atoms in component 1, when the GPE cannot be used.

\begin{figure*}[t]
\includegraphics[width=\linewidth]{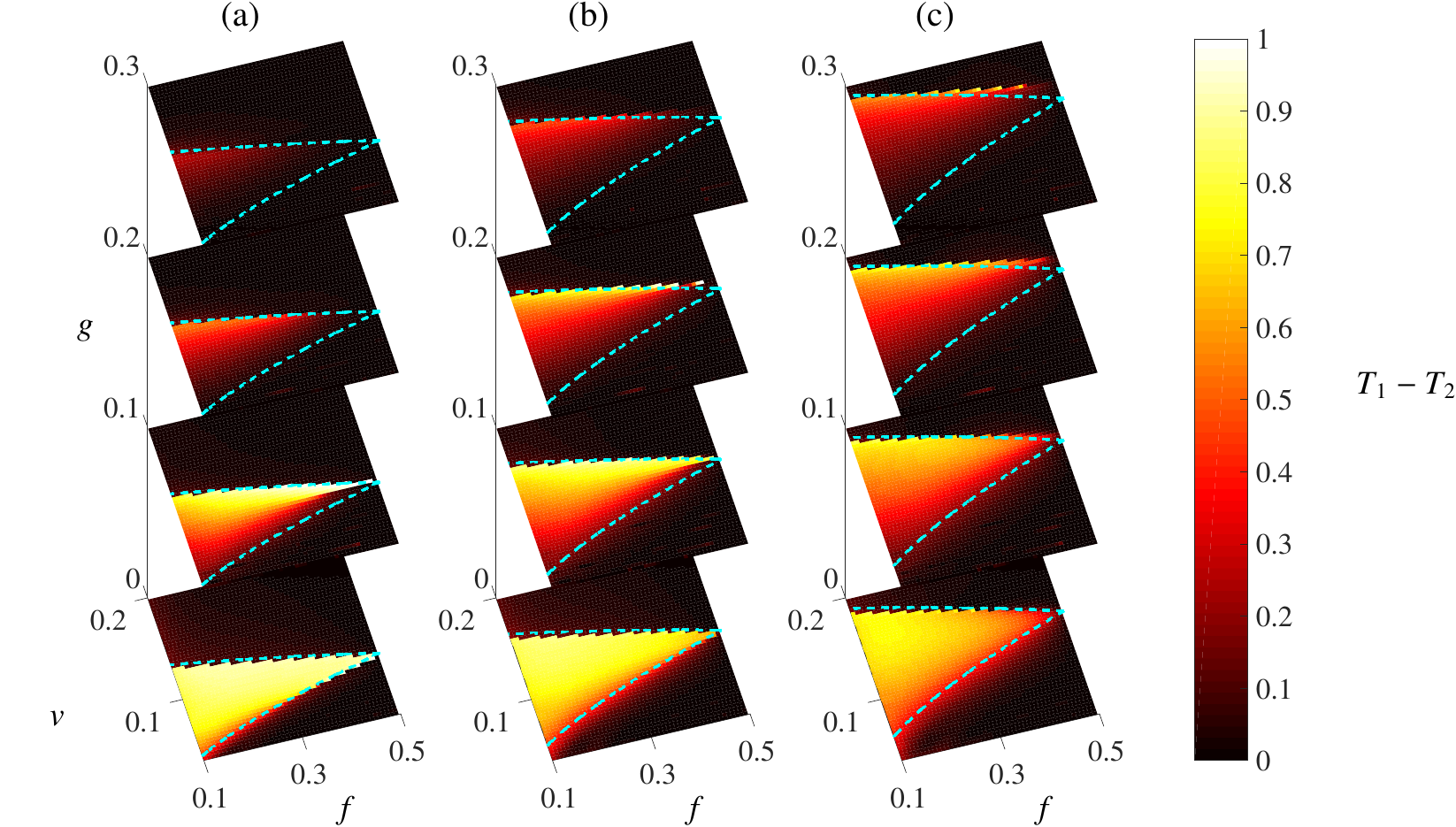}
\caption{Similar to Fig. \protect\ref{fig_ev3d}, but in the $(v,f)$
parameter plane at different values of $g$ and $\protect\varepsilon $. (a) $
\protect\varepsilon =0.04$, (b) $\protect\varepsilon =0.06$, and (c) $
\protect\varepsilon =0.08$.}
\label{fig_fv3d}
\end{figure*}

\section{Numerical results}

\begin{figure*}[t]
\includegraphics[width=\linewidth]{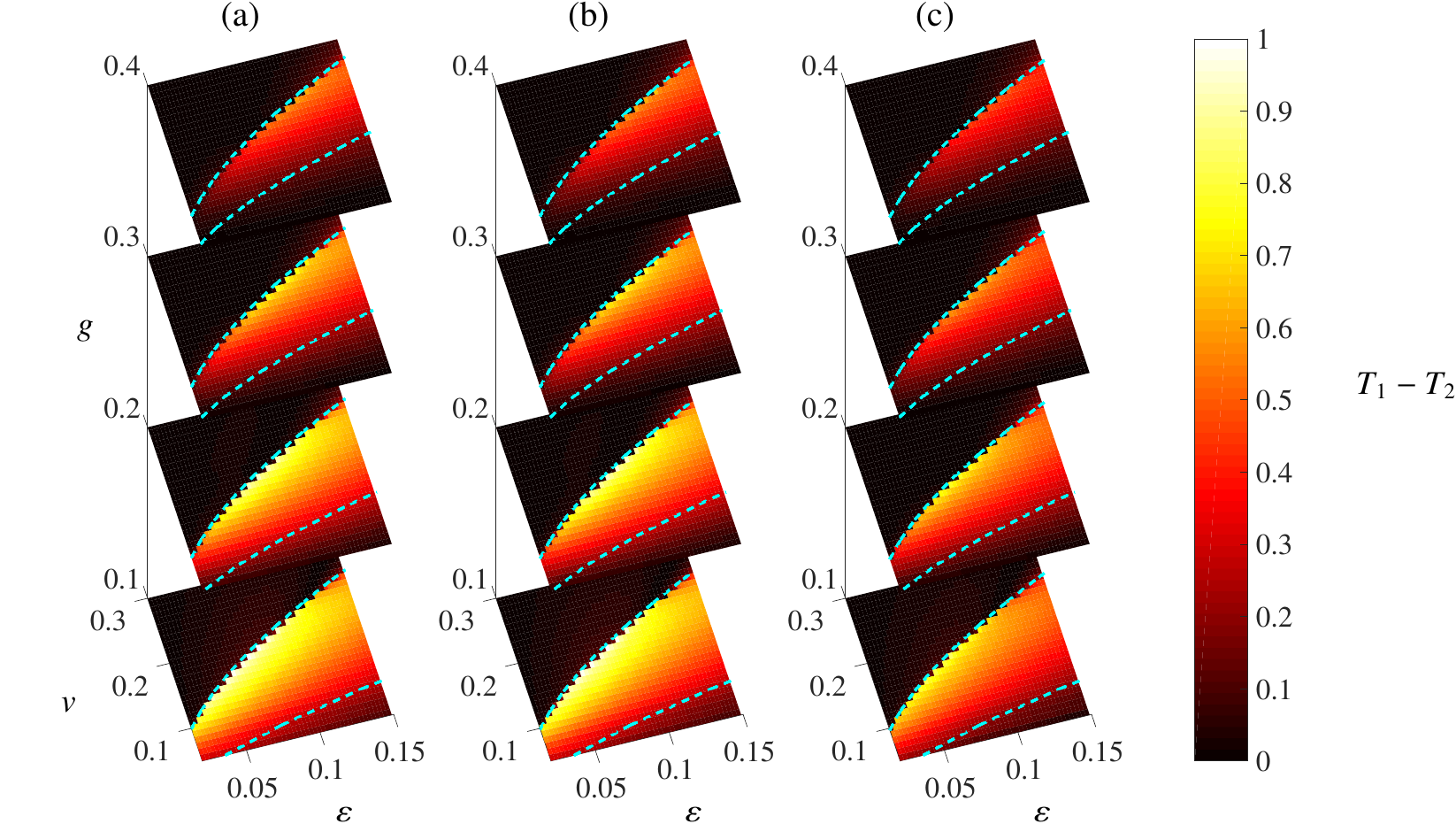}
\caption{The transmission difference between the two components, $T_{1}-T_{2}
$, as produced by simulations of Eqs. (\protect\ref{1}) and (\protect\ref{2}
), in the $(v,\protect\varepsilon )$ parameter space (where $v$ is the
collision velocity and $\protect\varepsilon $ the strength of the Gaussian
potential barrier) at different values of $g$ and $f$ (the cross-attraction
strength and scaled population of the first component, respectively). The
value of $g$ increases along the vertical axis. (a) $f=0.01$, (b) $f=0.02$,
and (c) $f=0.05$. The dashed lines display boundaries of the splitting
region, as analytically predicted by Eq.~(\protect\ref{g<1}) in the limit of
$f\ll 1$.}
\label{fig_ev3de}
\end{figure*}

\subsection{Details of the numerical approach}

In numerical simulations, we replace the ideal $\delta $-functional barrier
by a Gaussian one, $V(x)=\varepsilon \eta (x,\sigma )$, as defined in Eq.~(
\ref{sigma}). The intention here is to model an experimentally relevant
Gaussian-profile off-resonant sheet of light by the regularized version of
the $\delta $-function, with finite width $\sigma $. Except for Fig.~\ref
{fig_widths} and Fig.~\ref{fig_ev3ddf}, all numerical results presented in this paper were produced by
the Gaussian barrier with $\sigma =0.4$. This value of $\sigma $ is
reasonably small to adequately model the experimentally available barriers
\cite{si1}.

We numerically integrated Eqs.~(\ref{1}) and (\ref{2}), with the potential
barrier defined as per Eq.~(\ref{sigma}), by means of the well-known
Fourier-transform split-step method \cite{fssm1,fssm2}. We display typical
examples of collisions with the barrier, leading to either the splitting of
two-component solitons or their passage through the barrier, in Figs.~\ref
{traj_sep} and \ref{traj_nosep}, respectively. These two examples pertain to
slightly different collision velocities and otherwise identical parameters,
corresponding to situations close to the boundary between the separation and mutual
passage of the components. In Fig.~\ref{traj_sep} one can see that after the separation of the solitary waves into the smaller population transmitted component and the larger population reflected component the transmitted component has an increased width compared to before the collision. This is due to the diminished post-separation focussing term arising from the non-zero coupling to the larger population component ($g=0.2$ in Figs.~\ref{traj_sep} and \ref{traj_nosep}). Additionally, there is a small proportion of the smaller population component which is not transmitted in Fig.~\ref{traj_sep}. This would also lead to a larger width of the solitary wave due to decreased focussing by the non-linearity.

We quantify the transmission of the two components through the barrier by the coefficients
\begin{gather}
T_{1}=f^{~-1}\int_{0}^{\infty }\mathop{\mathrm{d}x}|\psi
_{1}(x,t=t_{f})|^{2},  \label{T1} \\
T_{2}=(1-f)^{-1}\int_{0}^{\infty }\mathop{\mathrm{d}x}|\psi
_{2}(x,t=t_{f})|^{2},  \label{T2}
\end{gather}%
which we compute at the \textquotedblleft final time\textquotedblright , $
t_{f}$. It is chosen to be $t_{f}\geq L/2v$ in the cases when the system
does not include an axial trapping potential, with $v$ being the velocity of
the incident soliton, and $L$ the size of the numerical spatial domain. We
set $L=160$, chosen so that, at the initial location ($x=-L/4$) and final
location of any transmitted component ($x=L/4$), the soliton components are
far separated from the splitting barrier. Note that the interaction with the
barrier decelerates the motion of any transmitted component, meaning that $
t_{f}$ must be increased accordingly. We have performed systematic
simulations to produce coefficients $T_{1,2}$ as functions of the four
control parameters, \textit{viz}., $v$, $f$, $\varepsilon $ and $g$. Note
that the repulsive barrier cannot trap any part of the wave functions,
meaning that in the absence (or negligibility) of the axial confinement the
reflection coefficients for the two components are $R_{1,2}=1-T_{1,2}$.

\subsection{Comparison of numerical results for the transmission with the
analytical predictions}

\subsubsection{Weak interspecies interaction}

In Figs.~\ref{fig_ev3d} and \ref{fig_fv3d} we compare the analytical
prediction, given by Eq.~(\ref{sep_condgll1}), with results of the
systematic simulations. In general, the agreement is good, provided that $g$
is small (a significant region for the value of $T_{1}-T_{2}=1$ is well
visible for up to about $g=0.1$, while the analytical approximation was
developed for $g=0$), and that $\varepsilon $ is also relatively small [up
to $\varepsilon \approx 0.1$, as expected for $f=0.3$ from Eq.~(\ref%
{applicability}), and seen in Fig.~\ref{fig_ev3d}].

In Figs.~\ref{fig_ev3d} and \ref{fig_fv3d}, we have mapped out the degree of
splitting, as produced by the simulations, in detail by plotting the
difference $T_{1}-T_{2}$ as a function of all the control parameters, $
\{\varepsilon ,v,g,f\}$. The same figures display boundaries (dashed lines)
between which the analytical result, given by Eq.~(\ref{sep_condgll1}),
predicts splitting to occur. To reiterate, the analytical consideration
implies that $T_{1}=1$ and $T_{2}=0$ in the interval of velocities of the
incident composite soliton given by Eq.~(\ref{sep_condgll1}), and, on the
other hand, $T_{1}=T_{2}$ outside the interval, where the incident soliton
does not split. It is clearly seen in Figs.~\ref{fig_ev3d} and \ref{fig_fv3d}
(as well as in Fig.~\ref{fig_ev3ddf}, which is produced below with an
effectively exact numerical implementation of a $\delta $-functional
barrier) that the analytical prediction gives a good indication of where
near-complete splitting occurs for $g\lesssim 1$, gradually deteriorating
with increasing $g$. This is explained, in particular, by the fact that, at
relatively large values of $g$, the attraction between the components
naturally tends to suppress the collision-induced splitting. It is also
generally true that, as the barrier area $\varepsilon $ increases (which,
for fixed width $\sigma $, effectively corresponds to increasing the
barrier's height), each component is only partly transmitted and partly
reflected in the simulations, i.e., $T_{1}$ and $T_{2}$ take intermediate
values between $0$ and $1$. The figures also corroborate the prediction of
Eq.~(\ref{sep_condgll1}), that the splitting region shrinks markedly as $f$
approaches $1/2$, i.e., the components of the incident composite solitons
become nearly equally populated.

\subsubsection{Strongly asymmetric nonlinearities}

We have also collected numerical results for the case of strong asymmetry
between the two components of the incident composite soliton, i.e.,
situations satisfying Eq.~(\ref{asymmetric_conditions}), in which case we
confine the consideration to $g^{\prime }=1$. These are collected in Fig.~
\ref{fig_ev3de}, which clearly demonstrates that the analytical prediction,
elaborated for this case in the form of Eq.~(\ref{g<1}), is quite accurate,
at least up to $f=0.05$, in broad intervals of values of $g$ and $
\varepsilon $. As in the case of small interspecies interactions, we cannot
expect perfect splitting (i.e.\ $T_{1}-T_{2}\neq 1$) for larger values of $g$
. This is codified in the case of strongly asymmetric nonlinearities in the
additional condition for separation given by Eq.~(\ref{Ecross_interval}),
although it produces only a qualitative indication.

Although we have found it simpler in our simulations to vary $f$ only, we
reiterate that it may not be experimentally practical to work with a very
small atomic population in one of the components, for reasons of imaging the
density profiles in the weakly populated component, and, generally speaking,
validity of the mean-field theory. On the other hand, simulations performed
for the alternative regime of $1\ll g\ll g^{\prime }$ with $f=1/2$ (not
shown here) yield qualitatively similar results. How best to fulfil Eq.~(\ref{asymmetric_conditions}) in a particular experimental configuration may
depend on what the available values of the scattering lengths are.

\subsection{Continuous variation of the interspecies interaction strength
with and without weak axial harmonic confinement}

In addition to considering the situation where a soliton moves with a given
velocity in free space, the simulations were also carried out for the
soliton beginning its motion from initial position $x_{0}$ on one side of an
external harmonic-oscillator potential,
\begin{equation}
U=\omega _{x}^{2}x^{2}/2.  \label{HO}
\end{equation}%
This setting implies that the soliton accelerates to incident velocity
\begin{equation}
v=\omega _{x}x_{0},  \label{omega x0}
\end{equation}%
when it hits the narrow barrier placed at $x=0$.

\begin{figure}[t]
\includegraphics[width=\linewidth,trim=1.5cm 0.0cm 0.0cm
0.5cm,clip=true]{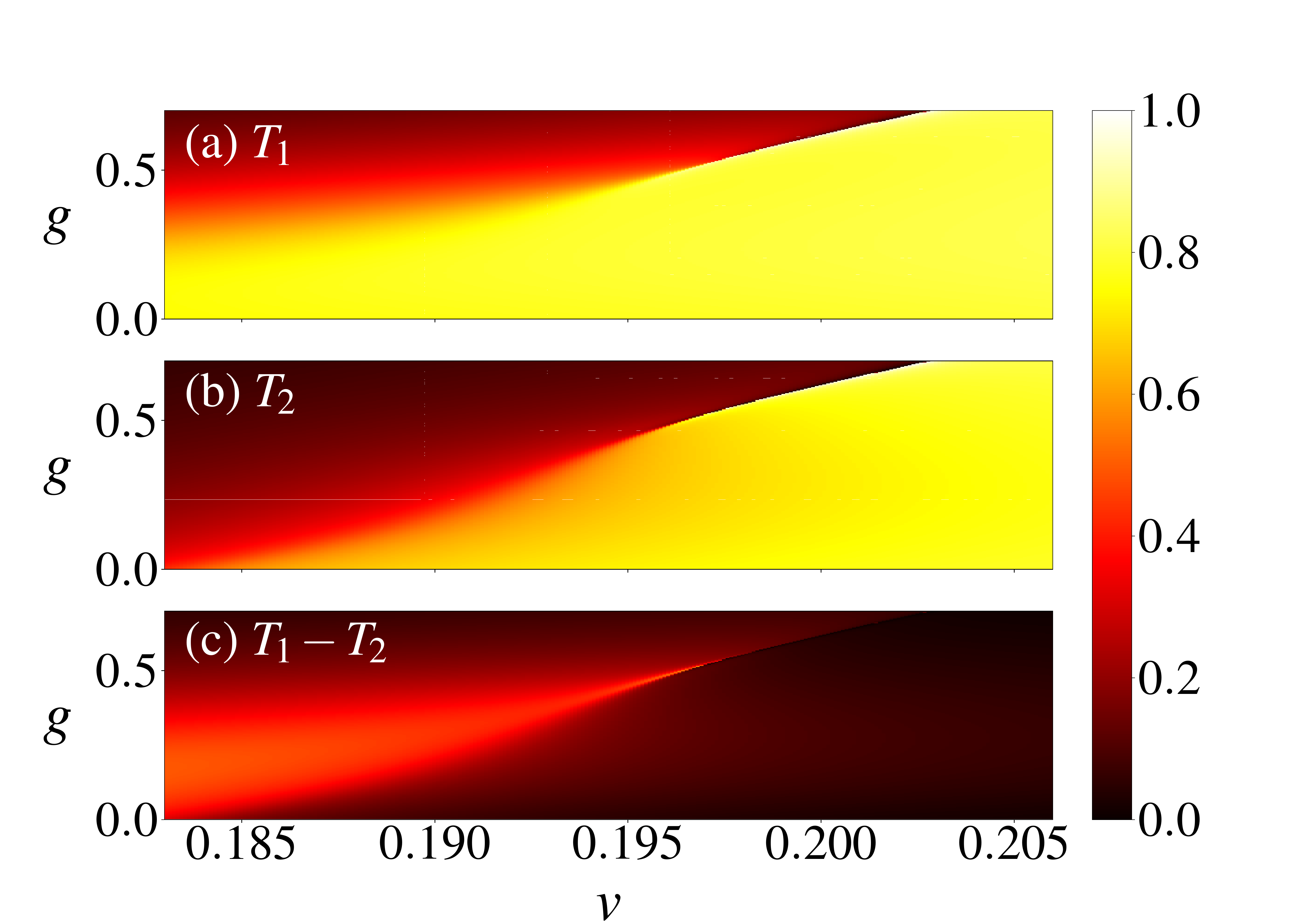}
\caption{Transmission coefficients of the two components, produced by the
simulations of Eqs. (\protect\ref{1}) and (\protect\ref{2}) for the
composite soliton, with the relative norm of the first component $f=0.3$,
incident on the splitter with strength $\protect\varepsilon =0.1$, The
results are displayed in the parameter plane of the collision velocity $v$
and interspecies attraction $g$. Panels (a), (b), and (c) display,
severally, the transmission coefficients of the first and second components,
and their difference.}
\label{fig_vg1}
\end{figure}

\begin{figure}[t]
\includegraphics[width=\linewidth,trim=1.5cm 0.0cm 0.0cm
0.5cm,clip=true]{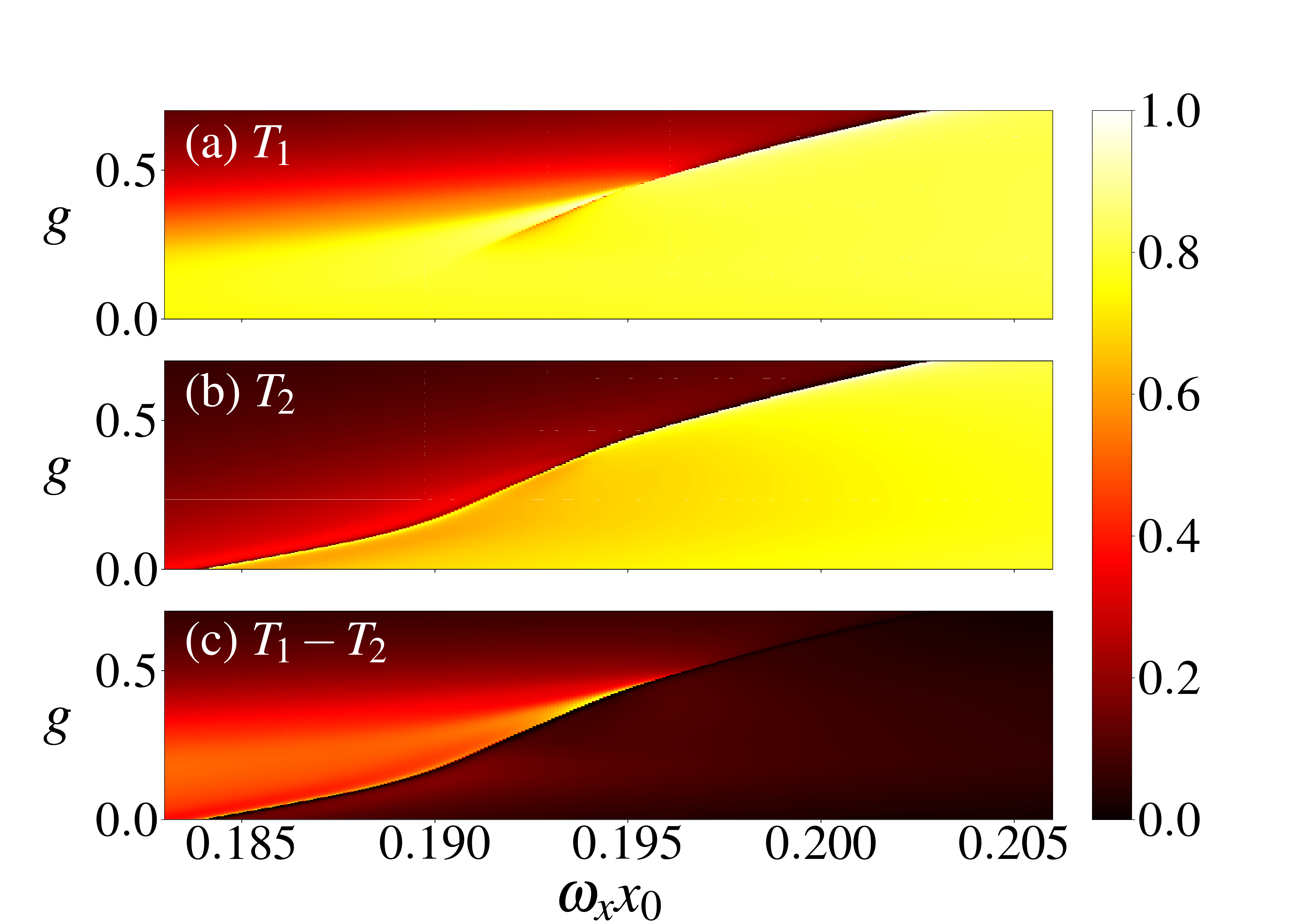}
\caption{The same as in Fig. \protect\ref{fig_vg1}, but in the case when the
collision velocity is given by Eq.~(\protect\ref{omega x0}) for the incident
soliton accelerated from initial position $x_{0}$ by the trapping potential (
\protect\ref{HO}).}
\label{fig_vgh1}
\end{figure}

Figure~\ref{fig_vg1} shows how the transmission in both components varies in
the $(v,g)$ parameter space for $\varepsilon =0.1$ and $f=0.3$ in the
free-space configuration (no axial trapping). In this case, Eq.~(\ref%
{sep_condgll1}) predicts that both components of the composite soliton pass
the barrier, without splitting, at $v>\sqrt{\varepsilon \left( 1-f\right) /2}
\approx 0.187$. The numerical findings collected in Fig.~\ref{fig_vg1}
support this prediction. This case can be compared to that when the axial
harmonic-oscillator confinement is present, as per Eq.~(\ref{HO}). Figure~
\ref{fig_vgh1} shows the results for an equivalent range of parameters with
the collision velocity given by Eq.~(\ref{omega x0}). In this figure, the
axial trapping produces a harder boundary in the $(v,g)$ parameter space,
separating the cases of component 2 being reflected and transmitted.

We display another aspect of the results, collected in Figs.~\ref{fig_vg1}
and Fig.~\ref{fig_vgh1}, in Fig.~\ref{fig_vgntandh1} by means of boundaries
between parameter regions where the second component is effectively
reflected or transmitted for different values of the barrier's strength, $
\varepsilon $, while fixing the proportion of the total population in this
component at $1-f=0.7$. We define this boundary by condition $T_{2}=0.5$.
Both sets of Figs.~\ref{fig_vg1} and \ref{fig_vgntandh1}(a), which pertain
to the soliton-barrier collision in free space, and Figs.~\ref{fig_vgh1} and 
\ref{fig_vgntandh1}(b), that display the numerical findings for the splitter
embedded in the external trapping potential [Eq.~(\ref{HO})], demonstrate
that making the attraction between the two components stronger, as
quantified by increasing the parameter $g$, while keeping other parameter
values fixed (relative population $f$ and barrier area $\varepsilon $),
leads to multiple transitions between positive and negative values of the
boundary's curvature.

\subsection{Internal excitations in past-collision solitons}

\begin{figure}[t]
\includegraphics[width=\linewidth,trim=0.5cm 0.5cm 2cm
0.8cm,clip=true]{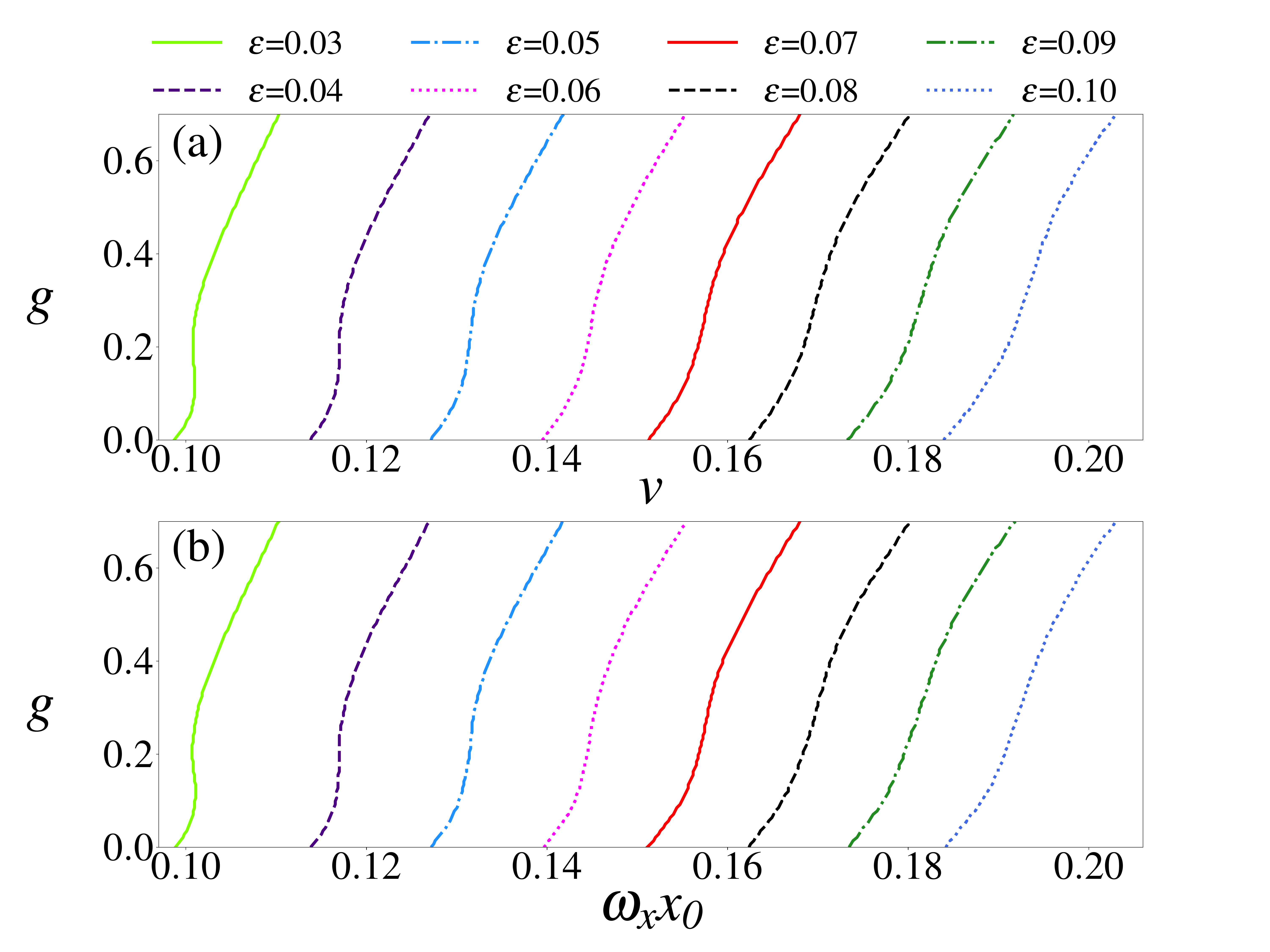}
\caption{Boundaries in the $(v,g)$ parameter plane (the collision velocity
and relative cross-attraction strength) between regions where the second
component of the incident composite soliton, with a fixed relative share of
the total norm, $1-f=0.7$, bounces (left of the boundary) or passes (right
of the boundary), for varying values of the barrier's strength, $\protect
\varepsilon $. (a) The case of the incident solitons arriving in free space
with velocity $v$; (b) for the soliton accelerated by the trapping potential
(\protect\ref{HO}) as per Eq.~(\protect\ref{omega x0}).}
\label{fig_vgntandh1}
\end{figure}

\begin{figure}[t]
\includegraphics[width=\linewidth,trim=0.5cm 0.0cm 4cm
3.0cm,clip=true]{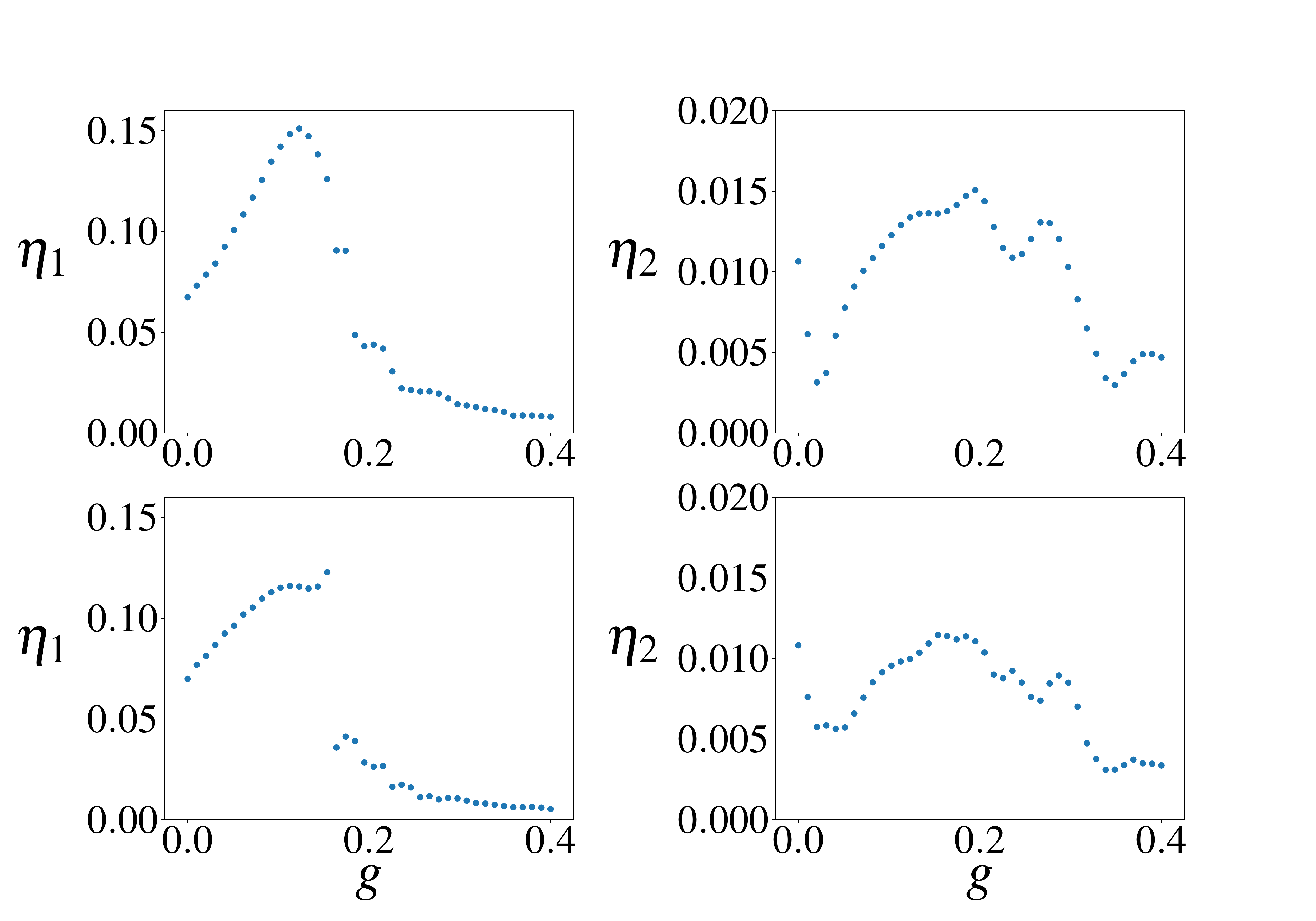}
\caption{The amount of the excitation of intrinsic vibrations in the passing
and rebounding solitons, in the first and second components, respectively,
defined as per Eq. (\protect\ref{def_eta}), for different values of coupling
strength $g$. The typical situation is displayed for other parameters fixed
as $f=0.3$, $v=0.112$, $\protect\varepsilon =0.04$. The top and bottom rows
summarize the results of the simulations performed, severally, in free space
and in the presence of the axial harmonic-oscillator trap. To clearly
observe post-collision internal oscillations in the solitons, a symmetric
spatial domain is used with the soliton initially placed at $|x_{0}|=200$,
in both cases.}
\label{fig_eta}
\end{figure}

An important aspect of the numerical results is the presence of intrinsic
excitations in the split and unsplit solitons after the interaction with the
barrier. In particular, for the use in interferometers the solitons should
keep a nearly-fundamental shape, without conspicuous internal vibrations. To
this end, we define the measure of the intrinsic excitation in the $j$-th
soliton ($j=1,2$) as
\begin{equation}
\eta _{j}=\frac{\max (n_{j})-\min (n_{j})}{\max (n_{j})+\min (n_{j})},
\label{def_eta}
\end{equation}%
where $n_{j}\equiv \left\vert \psi _{j}(x_c)\right\vert ^{2}$ is the density at the soliton's center, with the
maximum and minimum taken with respect to the evolution in time. It is shown
in Fig.~\ref{fig_eta} for different values of the interspecies coupling
strength, $g$, and $f=0.3$, $v=0.112$, $\varepsilon =0.04$, which adequately
represents a generic situation. In this case, the distance from the barrier,
$|x_{0}|$, was increased to $200$ in order to clearly observed the intrinsic
oscillations. In accordance with data displayed in Figs. \ref{fig_ev3d} and 
\ref{fig_fv3d}, the respective transmission coefficient of component 1, $
T_{1}$, is close to $1$ at $g=0$, decreasing to $0$ at $g=0.4$, while
component 2 bounces back, in agreement with the data collected in Fig. \ref{fig_vgntandh1}. Simultaneously, Fig. \ref{fig_eta} shows that the
excitation degree in the passing solitons (in component 1) first increases
from $\simeq 0.07$ to $\simeq 0.14$, and then decreases. The excitation in
the rebounding soliton (in component 2) follows a similar trend, but
remaining smaller, roughly, by an order of magnitude. These trends are
explained by an effect of increasing attraction between the components on
the excitation of the intrinsic vibrations in the solitons with the growth
of $g$, as well as by the effect of the varying shape of the interaction
products on the internal excitations in these products. A conclusion is
that, in the case of high-quality fission of the incident compound soliton
into the passing and rebounding ones, at $g$ sufficiently small, the
excitation effect remains weak. This conclusion is quite natural, as the
purity of the splitting deteriorates with the increase of $g$, which leads
to deformation of the splitting products, especially the passing soliton,
and the deformation excites the intrinsic vibrations.

It is relevant to stress that the excitation degree is much smaller than in
the previously studied case of the \textquotedblleft brute-force" splitting
of a single-component fundamental soliton by a strong barrier \cite{interf1,interf2,interf3,interf4,interf5,interf6,interf7,interf8,interf10,interf11}
. To address this point, we note that, in the case of the ideal splitting of
an incident fundamental soliton with amplitude $2A$ into two separated
fragments, each one, with center's coordinate $x_{0}$ and phase $\Phi _{0}$,
is naturally approximated by expression
\begin{equation}
\psi _{\mathrm{frag}}(x)=A\mathrm{\exp }\left( i\Phi _{0}\right) \mathrm{sech
}\left( a(x-x_{0}\right) ),a=2A,  \label{frag}
\end{equation}
cf. Eq. (\ref{psigll1and2}). Further evolution of the \textquotedblleft
half-soliton", initiated by this expression, can be produced by the exact
Satsuma-Yajima solution \cite{SY}, which is quite complicated. However,
states with the largest and smallest values of the density at the center,
which define the excitation measure (\ref{def_eta}), are ones with zero
chirp \cite{RMP}, which makes it possible to approximate them by ansatz (\ref{frag}) with independent values $A$ and $a$, subject to the conservation of
the norm,
\begin{equation}
\mathcal{N}\equiv \int_{-\infty }^{+\infty }\left\vert \psi (x)\right\vert
^{2}dx=2A^{2}a^{-1}  \label{N}
\end{equation}
(in the above analysis, normalization $\mathcal{N}=1$ is adopted). Further,
the substitution of the ansatz in the Hamiltonian of the ideal NLSE,
\begin{equation}
H_{\mathrm{single}}=\frac{1}{2}\int_{-\infty }^{+\infty }\left( \left\vert
\psi _{x}\right\vert ^{2}-|\psi |^{4}\right) dx  \label{single}
\end{equation}%
[cf. Eq. (\ref{H})], yields the corresponding value of the energy,
\begin{equation}
E_{\mathrm{single}}(A)=\frac{1}{3}\left( A^{2}a-2A^{4}a^{-1}\right) \equiv
\frac{1}{3}\left( 2\mathcal{N}^{-1}A^{4}-\mathcal{N}A^{2}\right) ,  \label{E}
\end{equation}%
where relation (\ref{N}) is used. Finally, equating energy (\ref{E}) for the
zero-chirp states realizing the states with maximal and minimal densities at
the center, the former one given by the split-soliton ansatz (\ref{frag}),
it is easy to find the relation between them, $\min (n)=(1/2)\max (n)$ (it
does not depend on $\mathcal{N}$). The respective value of the excitation
measure, as given by Eq. (\ref{def_eta}), is $\eta _{\mathrm{single}}=1/3$.
Thus, the above-mentioned typical values of $\eta _{j}$ for the two
components of the binary system are smaller than their counterparts in the
single-component model, in the case of the ideal splitting, by a factor $
\simeq 3-4,$ In the real single-component setting, numerical results yield
even larger values of the excitation measure, $\eta _{\mathrm{single}}\simeq
0.5~-0.6$ \cite{interf10}. Thus, the binary system provides, as expected,
essential suppression of detrimental effects of the post-collision intrinsic
excitation of the fragments.

\subsection{The effect of the finite barrier width}

\begin{figure}[t]
\includegraphics[width=\linewidth,trim=2cm 0.2cm 2cm
0.5cm,clip=true]{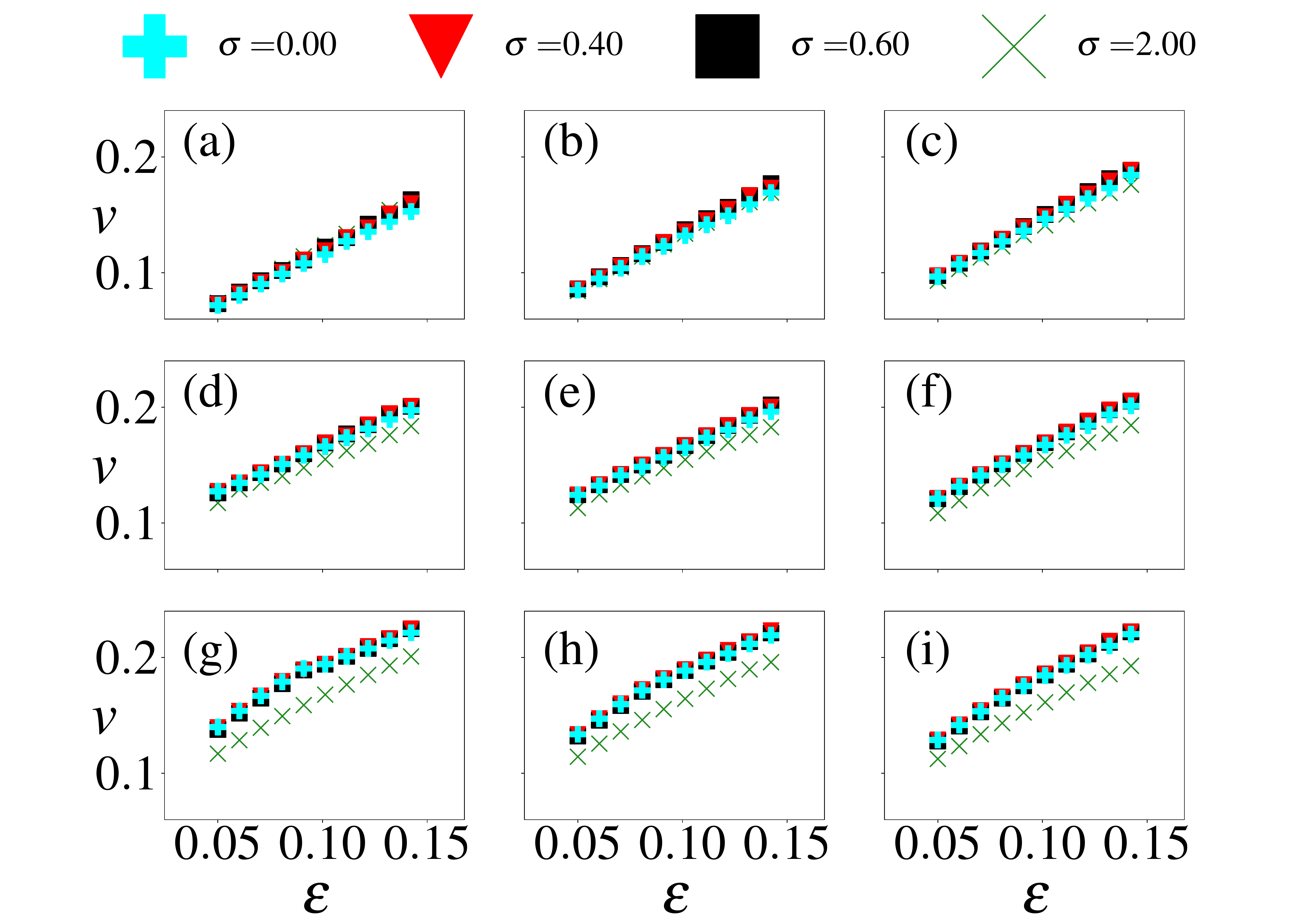}
\caption{Lines in parameter space ($\protect\varepsilon ,v$) at which the
transmission coefficient for the first component is $T_{1}=0.5$, for
different values of width $\protect\sigma $ of the Gaussian barrier (\protect
\ref{sigma}) ($\protect\sigma =0$ representing the ideal $\protect\delta $-
function). Other parameters are $f=0.2,g=0$ in (a), $f=0.3,g=0$ in (b), $
f=0.4,g=0$ in (c), $f=0.2,g=0.2$ in (d), $f=0.3,g=0.2$ in (e), $f=0.4,g=0.2$
(f), $f=0.2,g=0.4$ in (g), $f=0.3,g=0.4$ in (h), and $f=0.4,g=0.4$ in (i).}
\label{fig_widths}
\end{figure}

In Fig.~\ref{fig_widths} we show boundaries corresponding to $T_{1}=0.5$,
which separate the effective reflection and transmission of the first
component in the parameter space of collision velocity $v$ and barrier area $
\varepsilon $, for different fixed values of $f$, $g$, and the barrier width
$\sigma $ [see Eq.~(\ref{sigma})]. Note that we have obtained the results
for $\sigma =0$ by means of the numerical method outlined in Appendix~\ref{delta_barrier}, in which we represent the \textquotedblleft ideal" $\delta $
-functional barrier in Fourier space, and incorporate it in the split-step
simulation algorithm in the same step as the kinetic energy term [see Eq.~(
\ref{Fourier2})].

We choose parameter ranges in Fig.~\ref{fig_widths} so as to make them
representative for values used in Figs.~\ref{traj_sep}--\ref{fig_vgntandh1}.
From Fig.~\ref{fig_widths} one can see that the location in the $(\varepsilon ,v)$ parameter plane where $T_{1}=0.5$ is more sensitive to
width $\sigma $ of the barrier when $g$ is relatively large, and that for
the range of values of $f$ and $g$ considered in the analysis, the dynamics are, naturally, more sensitive to the variation of $g$ than the variation of $\sigma $. Increasing $\sigma $, while keeping other parameter values
constant, may cause the value of $v$ at which $T_{1}=0.5$ to become either
lower or higher, depending on the other parameters. In particular, a
conclusion is that, for $\sigma =0.4$, this value of $v$ is consistently
larger than that for the ideal $\delta $-function ($\sigma =0$), even if
this difference is never greater than $0.01$.

\section{Conclusions}

We have examined the transmission properties of two-component bright--bright
solitary waves colliding with a narrow potential barrier, and considered in
detail the effect of varying the barrier strength, incoming soliton
velocity, populations of its two components, scattering lengths. We have
carried this out with the main objective of identifying parameter regions in
which the incident composite soliton splits into its components so that one
is reflected and the other transmitted, which is an effect of major
importance to the design of matter-wave soliton interferometers. For small
values of the barrier strength $\varepsilon $, we have developed the
perturbation theory that effectively predicts the velocity interval in which
the splitting takes place for relatively weak interspecies interactions ($
g\rightarrow 0$), as well as for the case when the intraspecies interactions
are significant for one species only (which implies strongly different
populations of the two components, if the two intraspecies scattering
lengths are comparable). To obtain analytical estimates for these intervals,
we considered a $\delta $-functional barrier. In the numerical part of the
work, it was approximated by the corresponding Gaussian, an additional
parameter being its width $\sigma $. We have concluded that values of $
\sigma $ used in the simulations (such as $\sigma =0.4$, in scaled units),
produce results close to those that can be obtained with numerically exact
implementation of the $\delta $-function (in the Fourier-transform space).
By means of the comparison with numerical results, we have identified
parameter regions in which the perturbation theory accurately predicts the
outcomes of the collision of the incident composite soliton with the
splitting barrier. The numerical analysis was extended, in the parameter
space, beyond those regions, by increasing the strength of the interspecies
interaction, $g$, and varying the relative population in component 1, $f$.
The excitation of the intrinsic vibrations in the post-collision passing and
bounding solitons has also been studied, with a conclusion, that, in cases
of high-quality splitting, the excitations remain weak (much smaller than in
the previously studied single-component model).

Possible extensions of this work include numerical treatments of the system
with unequal intra-component scattering lengths and different atomic masses
in the two components (a heteronuclear binary BEC). Finally, it may also be
relevant to consider in detail the splitting by a localized nonlinear
potential, which is briefly addressed in Section \ref{Sec:ExtensionNonlinear}
of the Appendix.

Additional data related to the findings reported in this paper is made
available by source \cite{data}.

\begin{acknowledgments}
We would like to thank L. Tarruell for valuable comments on experimental data, and J. L. Helm for assistance with numerical techniques. B.A.M. appreciates support provided by the Durham University for collaborative work in the framework of the ``Structure" program, and C.L.G. is supported by the UK EPSRC. This work made use of the facilities of the Hamilton HPC Service of Durham University.
\end{acknowledgments}

\appendix

\section{Derivation of the transmission conditions\label{analysis}}

\subsection{The limit of the negligible interspecies interactions\label{lowganalysis}}

The center-of-mass kinetic energies of each component (essentially, half the
total mass of the soliton multiplied by the square of the velocity with
which it is moving) are written in our notation as [recall the unit energy
is $m(g_{11}N/\hbar )^{2}$ in physical units]
\begin{subequations}
\label{kinetic_energy}
\begin{align}
\left( E_{\mathrm{kin}}\right) _{1}=& fv^{2}/2,  \label{kinetic_energy1} \\
\left( E_{\mathrm{kin}}\right) _{2}=& (1-f)v^{2}/2,  \label{kinetic_energy2}
\end{align}%
and we determine the intraspecies interaction potential energies from terms
in the second line of Eq.~(\ref{H}):
\end{subequations}
\begin{subequations}
\label{interaction_energy}
\begin{align}
\left( E_{\mathrm{int}}\right) _{1}=& -f^{~3}/12, \\
\left( E_{\mathrm{int}}\right) _{2}=& -(1-f)^{3}\left( {g^{\prime }}\right)
^{2}/12.
\end{align}%
The potential energy of each component associated with the weak potential
barrier can be easily found in the framework of the perturbation theory
(assuming parameter $\varepsilon $ to be sufficiently small), which neglects the
deformation of the soliton under the action of the barrier potential \cite{RMP}:
\end{subequations}
\begin{subequations}
\label{Pot1and2}
\begin{align}
U_{1}(t)\equiv & \varepsilon \int_{-\infty }^{+\infty }\mathop{\mathrm{d}x}
\mathop{\delta (x)}|\psi _{1}(x,t)|^{2}  \notag \\
=& \frac{1}{4}\varepsilon f^{~2}\mathrm{sech}^{2}\left[ \frac{1}{2}f\left(
x_{0}+vt\right) \right] ,  \label{Pot1} \\
U_{2}(t)\equiv & \varepsilon \int_{-\infty }^{+\infty }\mathop{\mathrm{d}x}
\mathop{\delta (x)}|\psi _{2}(x,t)|^{2}  \notag \\
=& \frac{1}{4}\varepsilon (1-f)^{2}g^{\prime }\mathrm{sech}^{2}\left[ \frac{1
}{2}\left( 1-f\right) g^{\prime }\left( x_{0}+vt\right) \right] .
\label{Pot2}
\end{align}%
The perturbation theory that we have used applies (to each component)
provided that the magnitudes of the interaction potential [see Eqs.~(\ref{interaction_energy})] determine the peak values of the barrier-induced
energies, as given by Eqs.~(\ref{Pot1and2}) when $x_{0}+vt=0$. This yields,
eventually, the potential energy associated with the unperturbed component
solitons being located exactly on top of the barrier,
\end{subequations}
\begin{subequations}
\label{U12}
\begin{align}
(U_{\mathrm{max}})_{1}=& \varepsilon f^{2}/4, \\
(U_{\mathrm{max}})_{2}=& \varepsilon (1-f)^{2}g^{\prime }/4.
\end{align}

\begin{figure*}[t]
\includegraphics[width=\linewidth]{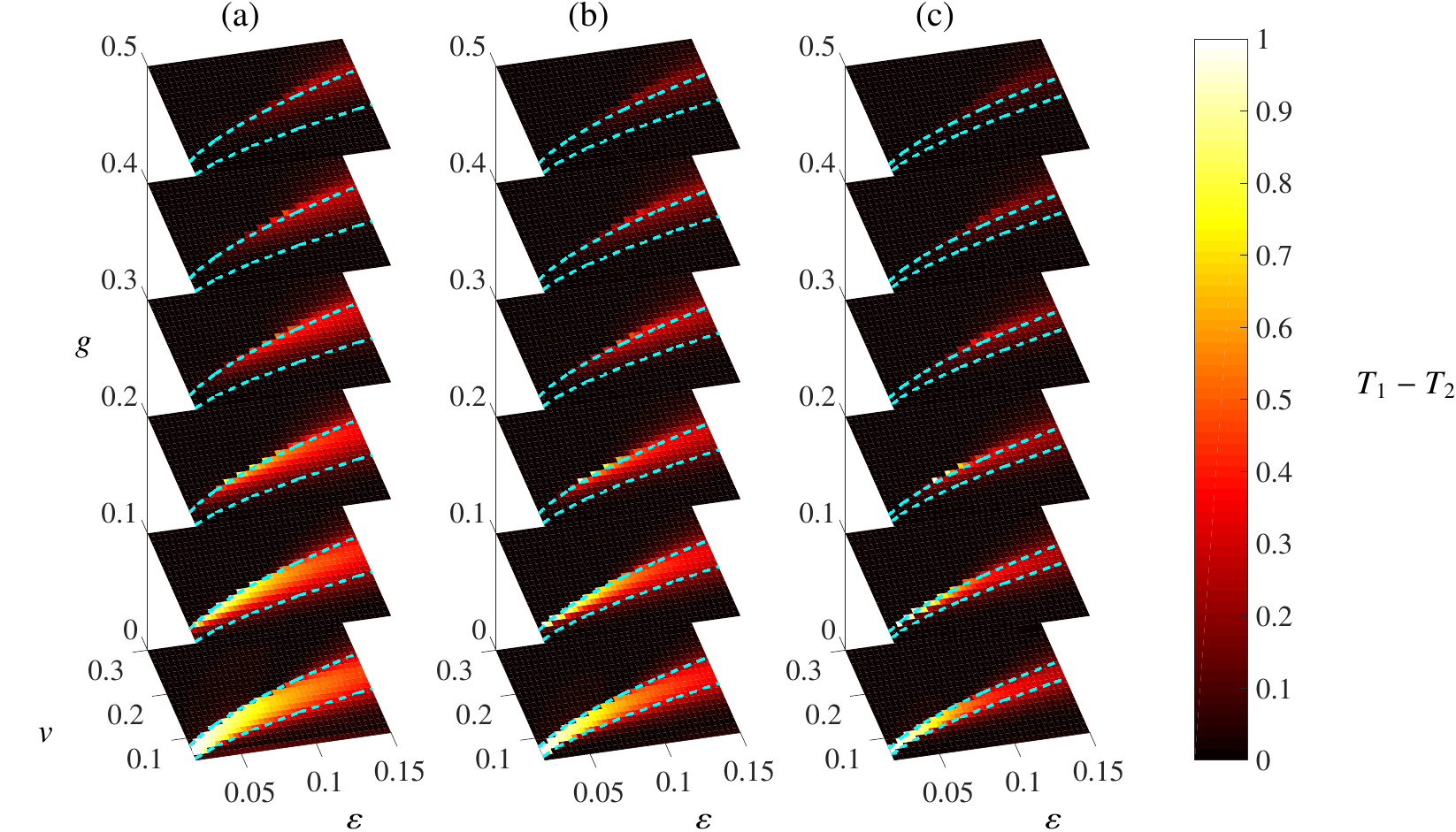}
\caption{The same as in Fig. \protect\ref{fig_ev3d}, but produced by the
numerical algorithm which implements the $\protect\delta $-function in the
Fourier space, as per Eq.~(\protect\ref{Fourier2}), instead of the Gaussian
barrier given by Eq.~(\protect\ref{sigma}) in the coordinate space.}
\label{fig_ev3ddf}
\end{figure*}

\subsection{Extension for the nonlinear splitter\label{Sec:ExtensionNonlinear}}

It is relatively straightforward to extend the theoretical treatment to the
case of a nonlinear splitter, as described in Refs.~\cite{HS} and \cite{Oleks}. In those works it took the form of a localized self-repulsive
nonlinearity (that can be created by a tightly focused laser beam which
locally applies by optical Feshbach resonance \cite{Tom}). The respectively
modified system of GPEs~(\ref{1and2}) is
\end{subequations}
\begin{subequations}
\begin{align}
i\frac{\partial \psi _{1}}{\partial t}& =\left[ -\frac{1}{2}\frac{\partial
^{2}}{\partial x^{2}}+\varepsilon _{\mathrm{nonlin}}\delta (x)\left\vert
\psi _{1}\right\vert ^{2}-|\psi _{1}|^{2}-g|\psi _{2}|^{2}\right] \psi _{1},
\label{nonlin1} \\
i\frac{\partial \psi _{2}}{\partial t}& =\left[ -\frac{1}{2}\frac{\partial
^{2}}{\partial x^{2}}+\varepsilon _{\mathrm{nonlin}}^{\prime }\delta
(x)\left\vert \psi _{2}\right\vert ^{2}-g^{\prime }|\psi _{2}|^{2}-g|\psi
_{1}|^{2}\right] \psi _{2},  \label{nonlin2}
\end{align}%
where positive $\varepsilon _{\mathrm{nonlin}}$ and $\varepsilon _{\mathrm{nonlin}}^{\prime }$ quantify the strengths of the nonlinear splitters,
which, in principle, may be different for the two atomic species.

Similarly to the linear case, when $g\ll 0$ one can determine velocity
intervals in which we predict splitting of an incident composite soliton
into a lighter transmitted soliton and a heavier reflected one:
\end{subequations}
\begin{equation}
\sqrt{\varepsilon _{\mathrm{nonlin}}f^{3}}/4<|v|<g^{\prime }\sqrt{\varepsilon _{\mathrm{nonlin}}^{\prime }(1-f)^{3}}/4.
\label{nonlin-interval}
\end{equation}%
Note that the form of this interval implicitly assumes $f^{3/2}\leq
(g^{\prime })^{3/2}\sqrt{\varepsilon _{\mathrm{nonlin}}^{\prime
}/\varepsilon _{\mathrm{nonlin}}}$, which is only fulfilled under the
condition that $f\leq (1-f)g^{\prime }$ when $g^{\prime }\sqrt{\varepsilon _{
\mathrm{nonlin}}^{\prime }/\varepsilon _{\mathrm{nonlin}}}\leq 1$. For
instance, in the case of $g^{\prime }=8$ and $f=13/15$ (with $\varepsilon _{
\mathrm{nonlin}}^{\prime }=\varepsilon _{\mathrm{nonlin}}$), the upper and
lower bounds of Eq.~(\ref{nonlin-interval}) need to be transposed.

One can readily determine nonlinear equivalents of Eq.~(\ref{U12}):
\begin{subequations}
\label{U12nonlinear}
\begin{align}
(U_{\mbox{\scriptsize max nonlin}})_{1}=& \varepsilon _{\mathrm{nonlin}
}f^{4}/32, \\
(U_{\mbox{\scriptsize max nonlin}})_{2}=& \varepsilon _{\mathrm{nonlin}
}^{\prime }(1-f)^{4}g^{\prime 2}/32,
\end{align}%
which, compared to Eq.~(\ref{interaction_energy}), reveal that the condition
for sufficiently small $\varepsilon _{\mathrm{nonlin}}$ is that is should be
significantly smaller than $8/3f$, and, similarly, $\varepsilon _{\mathrm{
nonlin}}^{\prime }$ should be significantly smaller than $8/3(1-f)$. If we
consider $\varepsilon _{\mathrm{nonlin}}$ and $\varepsilon _{\mathrm{nonlin}
}^{\prime }$ to be broadly similar in magnitude, this may be simplified to $
\{\varepsilon _{\mathrm{nonlin}},\varepsilon _{\mathrm{nonlin}}^{\prime
}\}\ll 8/3$.

Comparing Eq.~(\ref{nonlin-interval}) with Eq.~(\ref{sep_condgll1}), we see
that the nonlinear splitter manifests much stronger dependence on the
norm-distribution parameter $f=N_{1}/N$, as well as a stronger dependence on
the relative magnitude of the intraspecies scattering lengths, via $
g^{\prime }=a_{22}/a_{11}$.

\subsection{The limit of a strongly asymmetric two-component soliton}

\subsubsection{Determining the splitting interval \label{Sec:DeterminingSplitting}}

The kinetic energy, $\left( E_{\mathrm{kin}}\right) _{1}$, of component 1 is
given by Eq.~(\ref{kinetic_energy1}), while the height of the energy barrier
generated by the splitter, in similar fashion to Eq.~(\ref{Pot1}), may be
determined from Eq.~(\ref{N1}) as
\end{subequations}
\begin{equation}
(U_{\max })_{1}=\varepsilon A_{1}^{2}=\frac{\varepsilon f(1-f)g^{\prime }
\mathop{\Gamma \left( \alpha +1/2\right)}}{2\sqrt{\pi }\mathop{\Gamma(
\alpha)}}.  \label{max}
\end{equation}%
Combining these expressions within the energy condition for component 1 to
be transmitted through the barrier, $(E_{\mathrm{kin}})_{1}>(U_{\max })_{1}$
[cf.\ Eq.~(\ref{pass1})], yields the following result:
\begin{equation}
v^{2}>\frac{\varepsilon (1-f)g^{\prime }\mathop{\Gamma \left( \alpha
+1/2\right)}}{\sqrt{\pi }\mathop{\Gamma(\alpha)}}.
\end{equation}

\subsubsection{The interaction-energy condition \label{Sec:KineticEnergy}}

Strictly speaking, there is an additional condition necessary for the
complete collision-induced splitting in free space. The kinetic energy of
the transmitted component must exceed its binding energy in the composite
soliton, determined by the cross-attraction $E_{\mathrm{cross}}\equiv
-g\int_{-\infty }^{+\infty }\mathop{\mathrm{d}x}|\psi _{1}(x)|^{2}|\psi
_{2}(x)|^{2}$ [in this analysis we assume that the smallness of $\varepsilon
$ implies that the consideration of the energy described by Eq.~(\ref{max})
may be neglected altogether], otherwise component 1 will not become a free
soliton. Hence, making use of Eqs.~(\ref{psigll2}), (\ref{a}) and (\ref{N1}
), we obtain
\begin{align}
E_{\mathrm{cross}}=& -gA_{1}^{2}\frac{(1-f)^{2}g^{\prime }}{4}\int_{-\infty
}^{+\infty }\mathop{\mathrm{d}X}\left[ \mathrm{sech}\left( \frac{
[1-f]g^{\prime }X}{2}\right) \right] ^{2(\alpha +1)}  \notag \\
=& -gA_{1}^{2}\frac{(1-f)\sqrt{\pi }\mathop{\Gamma(\alpha +1)}}{2
\mathop{\Gamma(\alpha +3/2)}}  \notag \\
=& -\frac{\alpha gf(1-f)^{2}g^{\prime }}{2(2\alpha +1)}.  \label{Ecross}
\end{align}%
Substituting expressions (\ref{kinetic_energy}) and (\ref{Ecross}) in the
condition $(E_{\mathrm{kin}})_{1}>\left\vert E_{\mathrm{cross}}\right\vert $
yields the final constraint,
\begin{equation}
|v|>\sqrt{\frac{\alpha g(1-f)^{2}g^{\prime }}{\left( 2\alpha +1\right) }}.
\label{Ecross_interval}
\end{equation}

\section{Numerical simulations with regularized $\protect\delta $-functions
\label{delta_barrier}}

The scheme for handling the exact $\delta $-functional barrier in the
simulations is adapted from Ref. \cite{dfnumerics}. This incorporates the
Fourier transform of the $\delta $-function, $\hat{\delta}(k)$, into the
part of the split-step method which implements the kinetic-energy term. The
relevant expression for the split-step algorithm in the Fourier space is
then
\begin{equation}
\mathcal{F}[T+\varepsilon \delta (x)]=\frac{1}{2}k^{2}+\varepsilon \hat{
\delta}(k).  \label{Fourier1}
\end{equation}%
Due to the fact that one is conflating an analytical expression for the
Fourier transform and its discrete computational counterpart, one must be
careful while defining the periodic domain for the Fourier transform. To use
the discrete Fourier transform, in the numerical computations we choose the
domain as $-L/2\leq x<+L/2$, placing the $\delta $-function at the center.
The corresponding operator for the kinetic energy, combined with the energy
introduced by the $\delta $-function, is then written as
\begin{equation}
(M_{1})_{mn}=\mathcal{F}[T+\varepsilon \delta (x)]_{mn}=\frac{1}{2}
k^{2}\delta _{mn}+\frac{\varepsilon }{L},  \label{Fourier2}
\end{equation}%
where $k$ is defined as a discrete variable running between $-\pi /L$ and $
+\pi /L$ with $N$ entries, indexed by integers $\left( m,n\right) $, and $
\delta _{mn}$ is the Kronecker's delta. When using standard FFT routines in
the current context, they must be used in conjunction with two shifting
protocols (which shift the location of zero frequency to the centre of the
array) whenever they are applied, in order to run them in a way which is
consistent with the physically relevant boundary conditions.

Alternatively, one can use only one shifting protocol by accounting for a
phase offset in the resulting expression for the $\delta $-function in the
Fourier space. The expression for the sum of the kinetic energy with the
energy of the $\delta $-functional barrier is then
\begin{equation}
(M_{2})_{mn}=\mathcal{F}[T+\varepsilon \delta (x)]_{mn}=\frac{k^{2}}{2}
\delta _{mn}+\frac{\varepsilon }{L}\exp \left( \frac{iL}{2}%
[k_{m}-k_{n}]\right) ,  \label{Fourier3}
\end{equation}%
cf. Eq.~\ref{Fourier2}. Note that if we were only considering the kinetic
energy, this would make no difference, and it is in fact common practice to
only use one shifting protocol in this case.

To execute this step in the split-step algorithm, one must diagonalize
matrix $M_{1}$ or $M_{2}$ and combine the associated amount of shifts with
the Fourier transforms, as mentioned above. Note that the diagonalization
need only be done once, as it is constant throughout the simulations
(recalculation is required only if $\varepsilon $ or $L$ is altered). The
need, on a grid with $N$ spatial points, for $\left( N\times N\right) $
-dimensional matrix multiplications at each timestep, in order to implement
this method, increases the computational time. As a result, the resolution
of parameter space, as plotted in Fig.~\ref{fig_ev3ddf}, is reduced relative
to comparable plots presented above when using the Gaussian barrier.

Figure~\ref{fig_ev3ddf} shows a counterpart of Fig.~\ref{fig_ev3d}, produced
by the numerical scheme outlined in this appendix. Comparison of the plots
suggests that, for $g=0$, the results are quite similar, and the analytical
treatment gives a good indication of what to expect. Deviations of the
numerical results from the analytical approximation are primarily caused by
the nonlinearity and deformation of the solitons' shapes, rather than by the
deviation of the numerically approximated potential barrier (as long as it
is sufficiently narrow) from the ideal $\delta $-function.

\bibliographystyle{plain}
\bibliography{two_component_splitting}

\end{document}